\documentclass[12pt,preprint]{aastex}
\usepackage{emulateapj5}
\usepackage{epsfig}

\newcommand{\etal}{\mbox{et al.}}
\newcommand{\ergcms}{erg cm$^{-2}$ s$^{-1}$}
\newcommand{\ergsec}{erg s$^{-1}$}
\newcommand{\phcms}{photons cm$^{-2}$ s$^{-1}$}
\newcommand{\degree}{$^\circ$}

\newcommand{\chandra}{{\it Chandra}}
\newcommand{\rosat}{{\it ROSAT}}
\newcommand{\granat}{{\it GRANAT}}
\newcommand{\asca}{{\it ASCA}}
\newcommand{\einstein}{{\it Einstein}}
\newcommand{\bepposax}{{\it BeppoSAX}}

\newcommand{\sgrastar}{\mbox{Sgr A$^*$}}
\newcommand{\program}[1]{{\tt {#1}}}
\newcommand{\html}[1]{{\tt http://#1}}

\makeatletter
\newenvironment{inlinefigure}{%
\def\@captype{figure}%
\noindent\begin{minipage}{0.999\linewidth}\begin{center}}
{\end{center}\end{minipage}\smallskip}
\makeatother

\shortauthors{Muno \etal}
\shorttitle{Galactic Center {\it Chandra} Sources}

\begin{document}
\title{A Deep \chandra\ Catalog of X-ray Point Sources toward the Galactic 
Center}
\author{M. P. Muno,\altaffilmark{1} 
F. K. Baganoff,\altaffilmark{1} M. W. Bautz,\altaffilmark{1}
W. N. Brandt,\altaffilmark{2} P. S. Broos,\altaffilmark{2}
E. D. Feigelson,\altaffilmark{2} G. P. Garmire,\altaffilmark{2}
 M. R. Morris,\altaffilmark{3} G. R. Ricker,\altaffilmark{1}
and L. K. Townsley\altaffilmark{2}}

\altaffiltext{1}{Center for Space Research,
Massachusetts Institute of Technology, Cambridge, MA 02139;
muno@space.mit.edu, fkb@space.mit.edu}
\altaffiltext{2}{Department of Astronomy and Astrophysics, 
The Pennsylvania State University, University Park, PA 16802}
\altaffiltext{3}{Department of Physics and Astronomy, University of California,
Los Angeles, CA 90095}

\begin{abstract}
We present a catalog of 2357 point sources detected during 
590 ks of \chandra\ observations of the 17-by-17 arcminute field around 
\sgrastar.
This field encompasses a physical area of 40 by 40 pc at a distance of 8 kpc.
The completeness limit of the sample at the Galactic center is 
$10^{31}$ erg s$^{-1}$ (2.0--8.0 keV), while the detection limit is an order 
of magnitude lower. The 281 sources detected below 1.5 keV are mainly
in the foreground of the Galactic center, while comparisons to the 
\chandra\ deep fields at high Galactic latitudes suggest that only about 
100 of the observed sources are background AGN. The surface density of 
absorbed sources (not detected below 1.5~keV) 
falls off as $1/\theta$ away from \sgrastar, in agreement with the distribution
of stars in infrared surveys. This demonstrates the X-ray sources trace
the general stellar population at the Galactic center. Point sources brighter 
than our completeness limit produce 10\% of the flux previously attributed to 
diffuse emission. The $\log(N)-\log(S)$ distribution of the Galactic 
center sources is extremely steep (power-law slope $\alpha = 1.7$). If
this distribution extends down to a flux of $10^{-17}$ \ergcms\ 
($10^{29}$ \ergsec\ at 8~kpc, 2.0--8.0~keV) with the same slope,  
then point sources would account for all of the previously 
reported diffuse emission. However, there 
are numerous diffuse, filamentary structures in the field that also 
contribute to the total flux, so the 2.0--8.0 keV luminosity distribution 
must flatten between $10^{29} - 10^{31}$ \ergsec. 
Many types of stellar systems should be present in the field at the
luminosities to which we are sensitive. However, the spectra of more than 
half of the Galactic center sources are very hard, and can be
described by a power law ($E^{-\Gamma}$) with photon index $\Gamma < 1$. 
Such hard spectra have been seen previously only from magnetically accreting 
white dwarfs (polars and intermediate polars) and wind-accreting neutron 
stars (pulsars), suggesting that there are large numbers of these systems 
in our field.
\end{abstract}


\section{Introduction\label{sec:intro}}

The X-ray emission from galaxies is produced by a mixture of stellar sources 
at various phases of their life-cycles, diffuse interstellar plasma heated 
by supernovae and galactic collisions, and accretion onto super-massive black 
holes in galactic nuclei \citep[see][]{fab89}. With its 0\farcs5 angular
resolution, the {\it Chandra X-ray Observatory} is particularly 
well-suited to separating the diffuse and point-like components of this 
emission. \chandra\ observations can therefore be used to 
estimate more accurately the amount of hot, X-ray emitting interstellar 
matter in galaxies, and to trace the structures and star formation 
histories of galaxies using their stellar X-ray populations.

The total X-ray luminosity from galaxies that lack active nuclei is 
dominated by emission from neutron stars and black holes accreting from 
more ordinary stellar companions. 
These X-ray binaries have been observed in nearby galaxies with luminosities 
as low as $10^{36}$ \ergsec, allowing comparative studies of their luminosity 
distributions in galaxies with diverse star formation histories. 
For example, \citet{kil02} have established that galaxies 
with recent star formation contain relatively larger numbers of 
high-luminosity ($L_{\rm X} > 10^{38}$ \ergsec) X-ray sources than do 
non-starburst and elliptical galaxies. This suggests that 
many of the brightest X-ray binaries are fed by massive, short-lived 
stars. On the other hand, the X-ray luminosity functions of 
the bulges of nearby spiral galaxies appear to be flatter than those of 
their disks \citep{kon02b,sw02,tru02}. This indicates that the bright end 
of the luminosity distribution is dominated by old systems with 
low-mass companions when the pool of very young stars is smaller. 

All-sky surveys of our own Galaxy confirm that luminous 
($L_{\rm X} > 10^{35}$ \ergsec) X-ray binaries with high- and 
low-mass main-sequence companions reside in the Galactic 
disk and bulge, respectively (Grimm, Gilfanov, \& Sunyaev 2002)\nocite{gri02}.
However, \chandra\ observations
can reveal sources in our Galaxy as faint as $10^{31}$ \ergsec\
\citep[e.g.,][]{ebi01} at the Galactic center distance of 8~kpc 
\citep{mcn00}. Many additional types of stellar systems can be found 
down to this luminosity, (see Table~\ref{tab:ps} and 
references therein), which expands the possibility for using X-ray surveys
to study stellar populations.  X-rays from young stellar objects and 
cataclysmic variables can potentially be used to trace low-mass stars in 
regions of the Galactic disk and would complement current infrared surveys 
of luminous giants \citep{ug98,mez99}. Likewise, surveys of X-rays from 
O stars, Wolf-Rayet stars, and young neutron stars could be used to 
constrain the history of star formation within the last $10^8$ years. 
This would be particularly important 
in the inner tens of parsecs of the Galaxy, where it is uncertain how the 
large tidal forces and the milliGauss magnetic fields affect star 
formation, and where many traditional observational tracers of 
\begin{figure*}[t]
\centerline{\epsfig{file=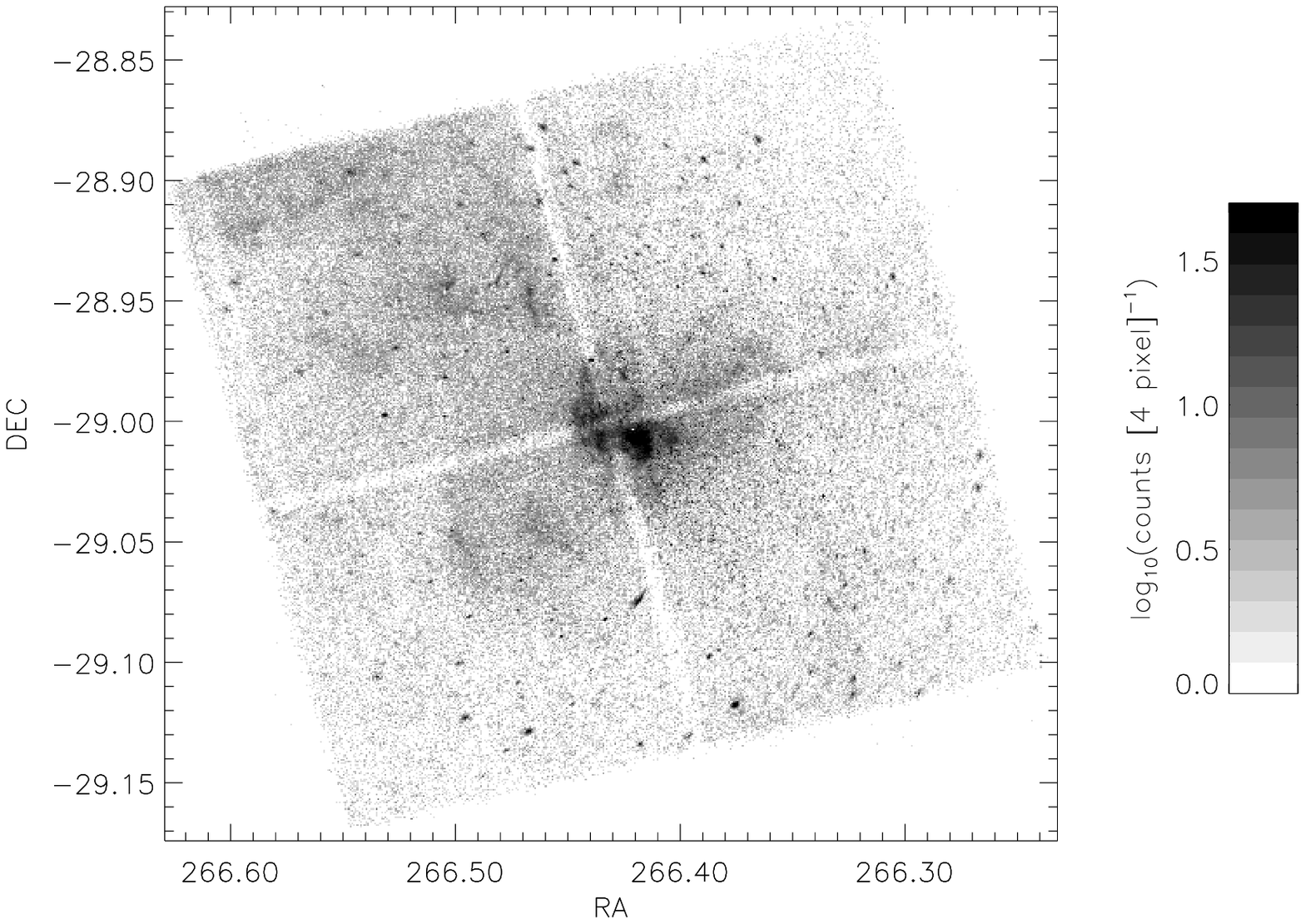,width=0.75\linewidth}}
\centerline{\epsfig{file=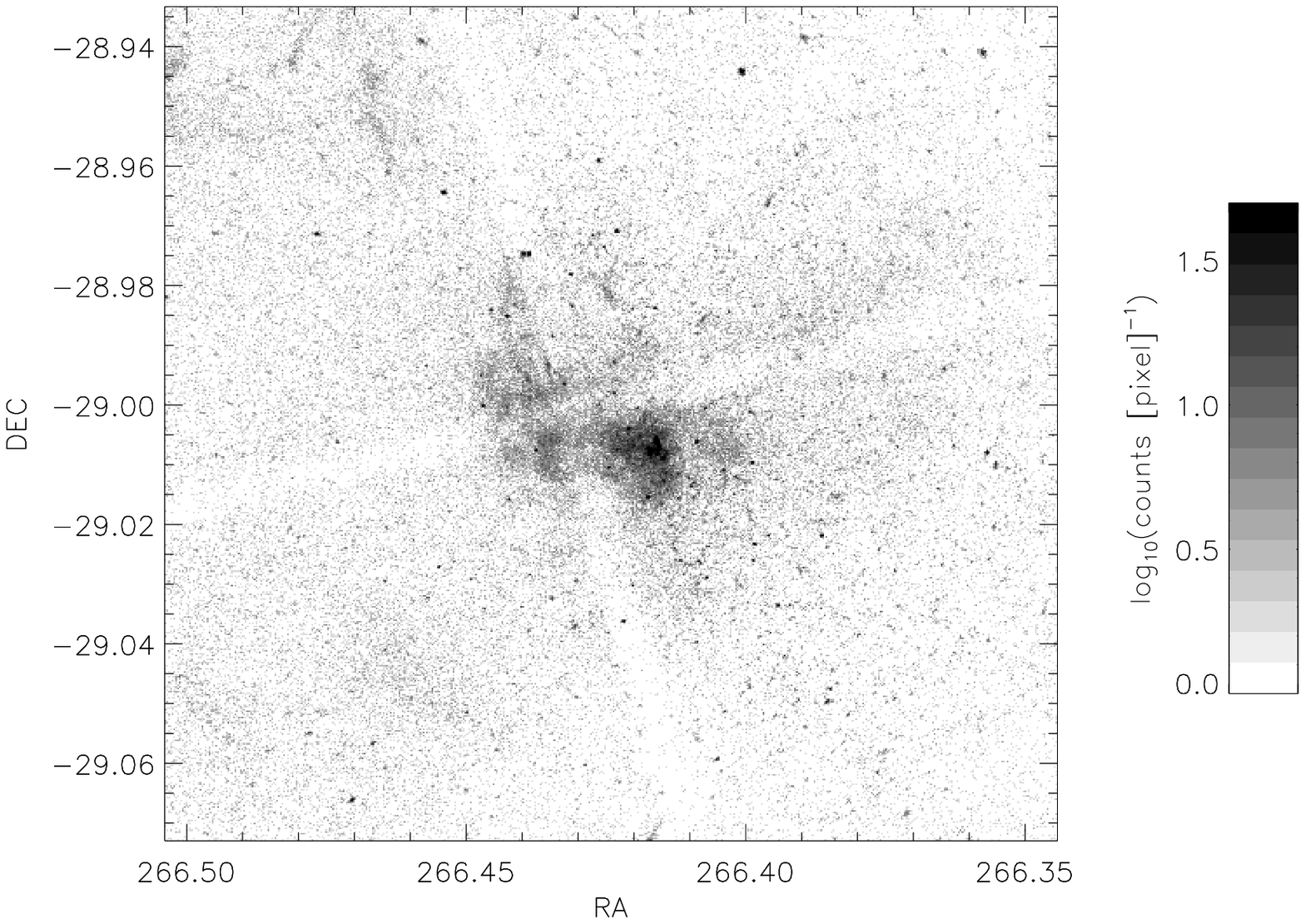,width=0.75\linewidth}}
\caption{Full-band images of the Galactic center field, 
uncorrected for variations in exposure time. The intensity is saturated
at high values and scaled logarithmically, as indicated by the color bar.
 {\it Top Panel:} 
The full 17\arcmin\ by 17\arcmin\ field, binned to cover 1024 by 1024 pixels.
{\it Bottom Panel}: The inner 8.5\arcmin\ by 8.5\arcmin\ field, at 
full resolution.}
\label{fig:rawimage}
\end{figure*}
star formation have been difficult to find 
\citep[][Mezger, Duschl, \& Zylka 1996]{mor93,sm96}\nocite{mdz96}. Faint 
X-ray 
sources have been used to study stellar populations only in small 
regions of the Galaxy, such as the Orion Nebula \citep[e.g.,][]{fei02b}, the 
Arches cluster \citep{yz02}, and several globular clusters 
\citep[e.g.,][]{pool02}. Large-area surveys with \einstein\ \citep{hg84}, 
\rosat\ \citep{mot97,mor01}, and \asca\ \citep{sug01} detected only the most 
luminous and nearby sources 
(see also Wang, Gotthelf, \& Lang 2002)\nocite{wang02}, 
while a recent \chandra\ survey at tens of 
degrees from the Galactic center found that the population of faint X-ray
sources was probably dominated by extra-galactic sources \citep{ebi01,ebi02}.
Further \chandra\ observations of dense stellar fields to understand the 
distribution of faint X-ray sources are therefore warranted.

If the number of faint X-ray sources toward the Galactic plane could be 
accurately counted, it also would be possible to constrain better the 
energetics of the diffuse emission from the Galactic 
ridge \citep{koy86b, yam96, ms93, sug01}. 
This diffuse emission appears to be a combination of relatively cool 
($kT \sim 0.3$~keV) thermal emission, and hotter ($kT>7$~keV) emission 
that may extend all the way to MeV energies 
\citep[Koyama, Ikeuchi, \& Tomisaka 1986a,][]{koy86b, kan97, yam97, ski97}. 
The cool, thermal component 
can be explained as the integrated emission from unresolved supernova 
remnants \citep{koy86a}. However, the temperature of 
the hot emission is much higher than that observed from supernova shocks, 
and it is too high for the plasma thought to produce it to be gravitationally 
bound to the Galactic disk \citep{wor82,koy86b}. If the plasma is 
unbound, the energy input required to sustain this hard Galactic ridge 
emission is approximately $10^{42}$ \ergsec, equivalent to the kinetic energy 
of one supernova occurring every 30 years \citep{vm98}. This 
input would have to be provided by exotic processes, such as cosmic-ray 
interactions with the 
ISM \citep[][Tanaka, Miyaji, \& Hasinger 1999]{vm98}\nocite{tmh99} 
or magnetic reconnection driven by turbulence in the ISM \citep{tan99}. 
No candidate population of point
sources has yet been identified that could significantly lessen the 
energetic requirements on the plasma. \chandra\ observations at 
$l = 28^\circ$ and $b = 0.2^{\circ}$ indicate that only 10\% of the hard 
Galactic ridge emission can be accounted for by 
X-ray point sources brighter than $10^{31}$ \ergsec\ \citep{ebi01}.
However, fainter sources could still contribute significantly to the diffuse 
emission (see Table~\ref{tab:ps}), if they are present in large numbers.

The nucleus of our Galaxy is an ideal location to explore these
topics, since both the stellar density \citep{mdz96} and the 
surface brightness of diffuse X-ray emission \citep{kan97,koy96,sm99} 
increase dramatically there. The Galactic center has been the object of 
observations with \einstein\ \citep{wat81}, 
\granat\ (Pavlinsky, Grebenev, \& Sunyaev 1994)\nocite{pav94}, 
\rosat\ \citep[][Sidoli, Belloni, \& Mereghetti 2001]{pt94}\nocite{sbm01}, 
\bepposax\ \citep{sid99}, and \asca\ \citep{sak02}, all of which revealed 
several bright ($>10^{35}$ \ergsec) point sources and emission from 
the Sgr A complex. \chandra\ was the first instrument to resolve the X-ray 
emission from the accreting black hole \sgrastar\ from the 
surrounding early-type stars, the remnant of 
a $10^{52}$ erg explosion \citep[Sgr A East;][]{mae02}, and numerous 
filamentary features \citep{bag01,bag02}. Over 150 point sources were 
also detected in the 17\arcmin\ by 17\arcmin\ field, down to a limiting flux 
of $2\times10^{-14}$ \ergcms\ 
\citep[$L_{\rm x} = 2\times10^{32}$ \ergsec\ at 8 kpc;][]{bag02}. 

Recent observations have increased the \chandra\ exposure of the 20 pc
around \sgrastar\ by a factor of 6, to 626~ks. In this paper, we 
present a catalog of 2357 X-ray point sources detected in this field.
In Section~\ref{sec:obs}, we describe the observations and
our source detection method (Section~\ref{sec:det}), 
our technique for computing the photometry of the sources 
(Section~\ref{sec:phot}), and our estimates of the 
completeness of the survey (Section~\ref{sec:area}). 
In Section~\ref{sec:res}, we report the spatial 
distribution (Section~\ref{sec:spat}), flux distribution 
(Section~\ref{sec:flux}), confusion limit (Section~\ref{sec:conf}), 
and spectral properties (Sections~\ref{sec:hr} and \ref{sec:spec}) 
of the sources. In Section~\ref{sec:dis},
we discuss how the numbers of sources at the Galactic center compare to 
populations elsewhere in the 
Galactic disk and to background AGN (Section~\ref{sec:dis:num}),
the contribution of point sources to the diffuse emission from the Galactic
center (Section~\ref{sec:dis:dif}), and the possible nature of the point 
sources (Section~\ref{sec:dis:num}).

\section{Observations and Data Analysis\label{sec:obs}}

Twelve separate pointings toward the Galactic center have been carried out 
using the Advanced CCD Imaging Spectrometer imaging array (ACIS-I) aboard 
the {\it Chandra X-ray Observatory} \citep{gar02}, 
in order to monitor \sgrastar\ (Table~\ref{tab:obs}).
The ACIS-I is a set of four 1024-by-1024 pixel CCDs, covering
a field of view of 17\arcmin\ by 17\arcmin. When placed on-axis at the focal
plane of the grazing-incidence X-ray mirrors, the imaging resolution 
is determined by the pixel size of the CCDs, 0\farcs492. The CCDs also 
measure the energy of incident photons, with a resolution of 50-300 eV 
(depending on photon energy and distance from the read-out node), within a 
calibrated energy band of 0.5--8~keV.

We reduced the data starting with the level 1 event files provided by 
the Chandra X-ray Center (CXC). We first removed the pixel randomization 
applied by the default processing software. We then corrected the pulse 
heights of each event for the position-dependent
charge-transfer inefficiency caused by radiation damage early in the 
mission, using software provided by \citet{tow00}.  We excluded most events
flagged as possible background, but left in possible cosmic ray
afterglows because they are difficult to distinguish from genuine X-rays 
from the strong diffuse emission and numerous point 
sources in the field. We applied the standard ASCA grade filters to the 
events, as well as the good-time filters supplied by the CXC. We examined each
observation for background flares, and removed intervals of strong flaring
from ObsID 0242 (10 ks), ObsID 2943 (3 ks), and ObsIDs 2953, 3392, and 
3393 (each $< 1$~ks). The final exposure time was 626 ks. 
Finally, we applied a correction to the absolute
astrometry of each pointing, using three Tycho sources detected strongly
in each \chandra\ observation \citep[compare][]{bag02}. The astrometric 
accuracy of our final pointing solution is better than 0\farcs3, although
the accuracy of positions derived for individual sources decreases 
significantly far from the aim point (see Section~\ref{sec:det}). 

In order to produce a single composite image, the sky coordinates of the 
events from each observation were re-projected to the tangent plane at the 
radio position of \sgrastar\ 
\begin{figure*}[t]
\centerline{\epsfig{file=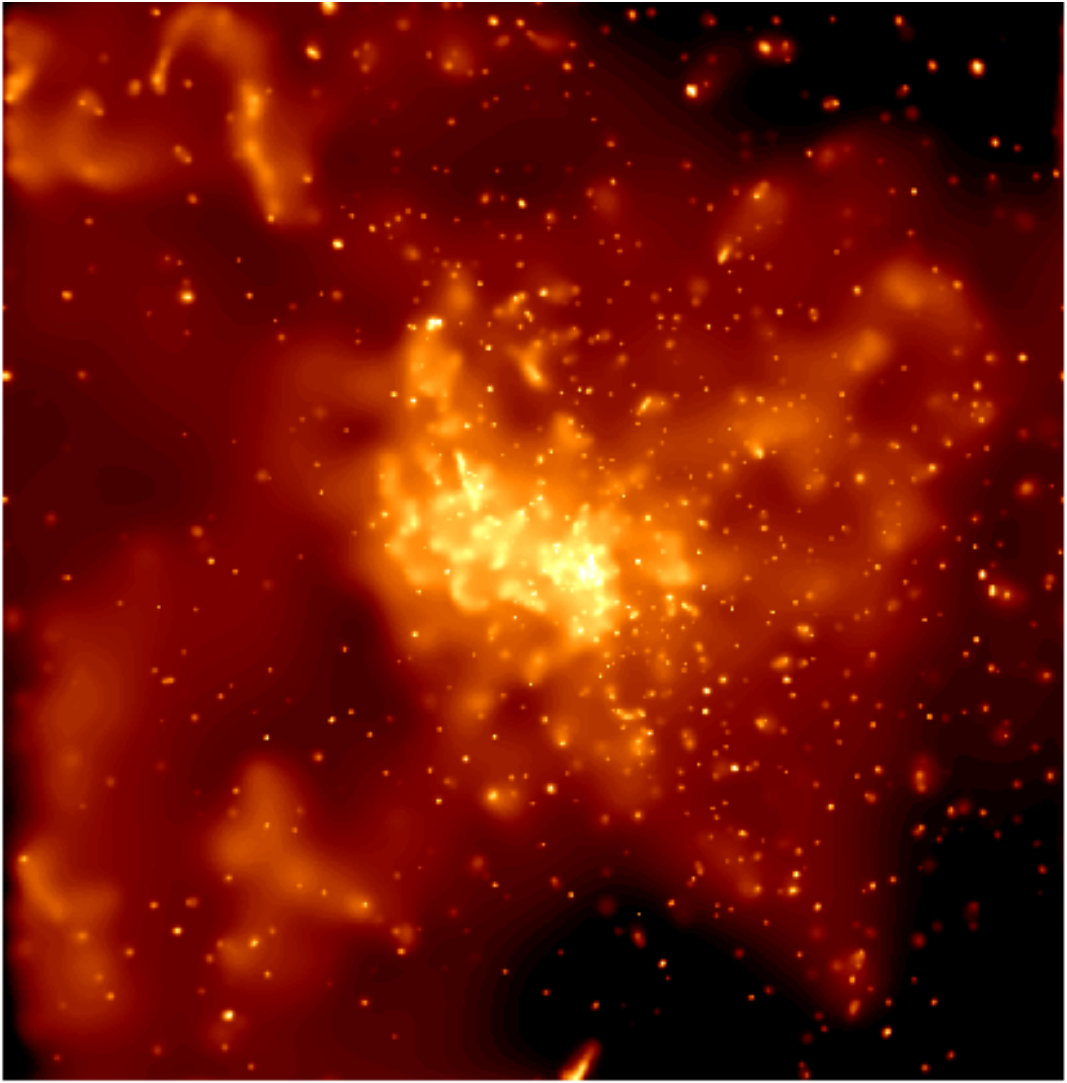,width=\linewidth}}
\caption{Full-band image of the inner 8.5\arcmin\ by 8.5\arcmin\ field 
around Sgr A*, as in Figure~1b. The image has been corrected for variations in exposure
due to bad columns and chip gaps, and has been adaptively smoothed to 
allow point sources and diffuse emission to be distinguished more easily.
The color scale is logarithmic.
}
\label{fig:image}
\end{figure*}
\citep[17h 45m 40.0409(9)s, -29\degree 00\arcmin 28\farcs118(12);][]{rei99}, 
and the event lists were combined. Images of the composite event list 
are displayed in Figure~\ref{fig:rawimage}. We 
excluded ObsID 1561a 
(2000 October 26; Table~\ref{tab:obs}),
because it contained a bright transient with a large dust scattering
halo and instrumental readout streak.\footnote{The transient is probably
GRS~1741.9$-$2853, which we place at J2000 17h 45m 2s, 
-28\degree 54\arcmin 51\arcsec\ with an uncertainty of 5\arcsec; 
compare Pavlinsky et al.\ (1994).} Exposure maps were created 
assuming a monochromatic incident spectrum with a photon energy of 3~keV, 
which is the 
approximate energy at which the largest number of photons are detected. 
In Figure~\ref{fig:image}, we display a version of the inner part of the 
image that has been corrected for exposure variations over the field, and 
then adaptively smoothed using the program \program{csmooth}. 

\subsection{Source Detection\label{sec:det}}

\begin{figure*}[t]
\centerline{\epsfig{file=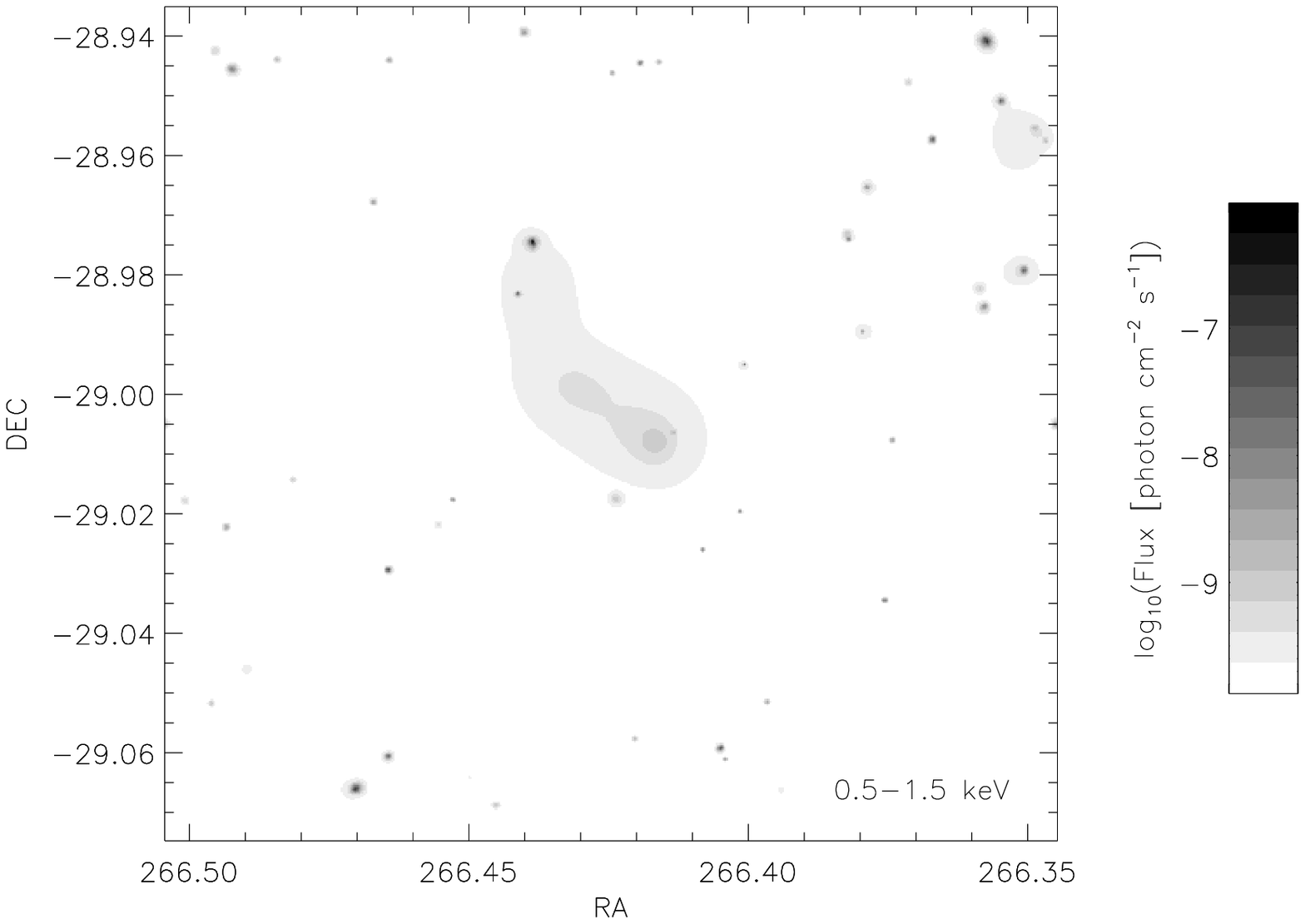,width=0.65\linewidth}}
\centerline{\epsfig{file=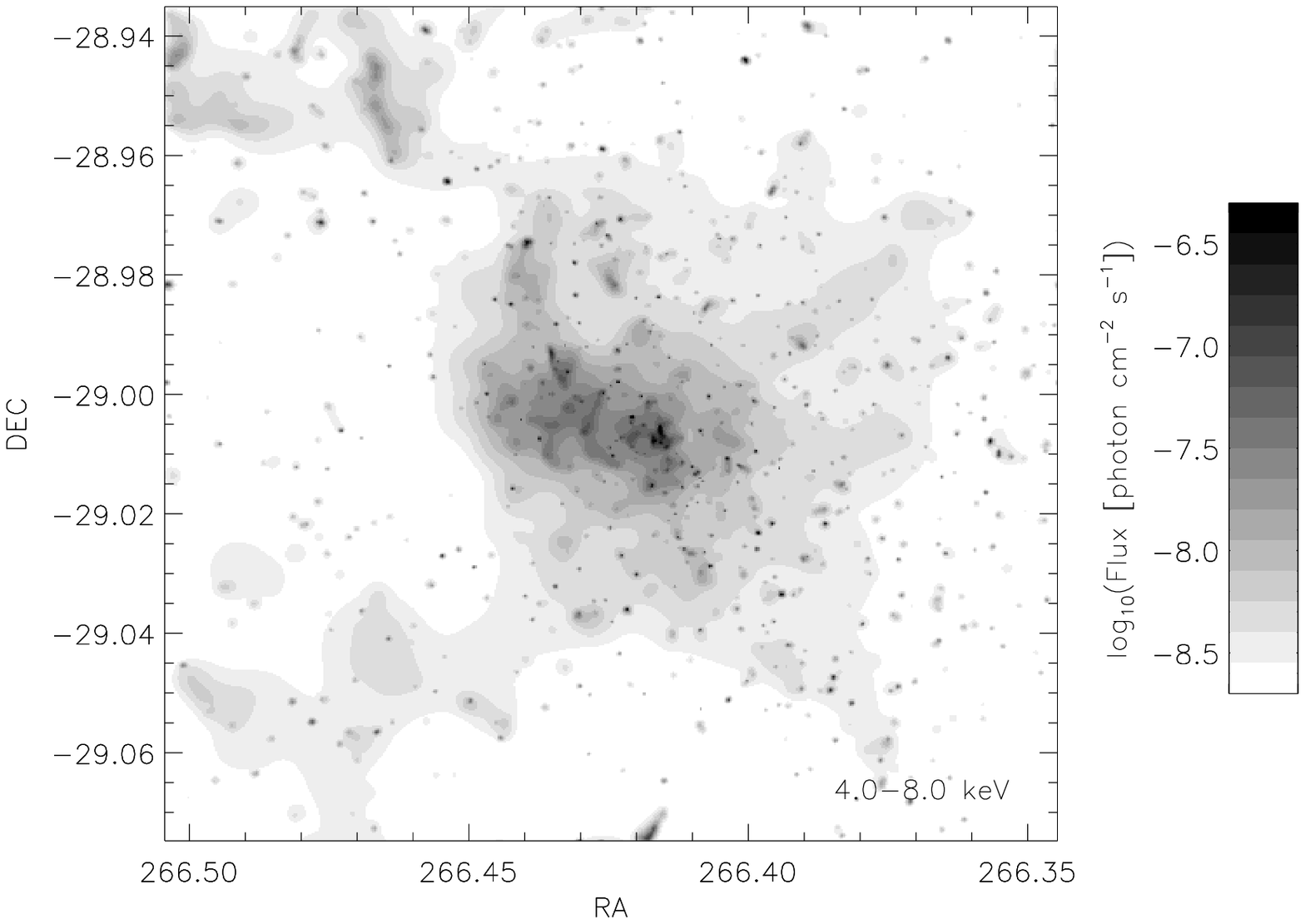,width=0.65\linewidth}}
\caption{Same as Figure~2, for the 0.5--1.5~keV and the
4--8~keV energy bands.}
\label{fig:energy_images}
\end{figure*}

We used the combined event list (excluding ObsID 1561a) to search for point 
sources in three
energy bands: a full band extending from 0.5--8~keV, a soft band from
0.5--1.5~keV, and a hard band from 4--8~keV. Smoothed images of the soft and
hard energy bands are shown in Figure~\ref{fig:energy_images}. For the 
purposes of 
source detection only, we removed events that had been flagged as 
possible cosmic ray afterglows. We employed
the routine \program{wavdetect} \citep{fre02}, 
using the default ``Mexican Hat'' wavelet. We searched
a series of three images using sequences of wavelet scales that increased 
by a factor of $\sqrt{2}$: a central, un-binned image of 
8.5\arcmin\ by 8.5\arcmin\ searched from scales 1--4, an image binned by a 
factor of two to cover 17\arcmin\ by 17\arcmin\ searched from scales 1--8, and 
an image binned by a factor of four to cover the entire field 
searched from scales 1--16 (since 
observations were taken with slightly different aim points and roll angles). 
Each image was designed to be most sensitive to point sources located
at successively larger field offset angles, since the 90\% encircled energy 
contour of the point-spread function (PSF) at 4.5~keV 
grows from 2\arcsec\ near the aim point, to 15\arcsec\ at 10\arcmin\ from
the aim point (see the \chandra\ 
Proposers' Observatory Guide). 

We used a sensitivity threshold of $10^{-7}$, which was determined via
Monte-Carlo simulations to be the chance of 
detecting a spurious source
per pixel if the local background is spatially uniform \citep{fre02}. 
Since the strongly varying diffuse emission in our field invalidates this 
assumption, we can not be sure of the true significance that this threshold 
represents (compare Section~\ref{sec:phot}). 
However, this threshold is conservative
\citep[compare][]{bra01,fei02b}, so there should be 
few spurious sources in our sample.\footnote{In addition, the extra sources 
that would be detected with looser detection thresholds would be too faint 
for meaningful spectral analyses (Section~\ref{sec:hr}), 
and the uncertainties on 
their fluxes would be too large to use them to constrain the 
$\log(N)-\log(S)$ distribution (Section~\ref{sec:flux}).} 
We combined the 
source lists generated from each image, including only the sources from the 
images with higher resolution in the regions that overlapped. When combining 
the candidate sources from the separate energy bands, we gave priority to the
source positions determined in the full band, and considered two sources 
to be the same if they were separated by less than one-half the 90\% 
encircled energy radius of the PSF at 
\begin{figure*}[t]
\centerline{\epsfig{file=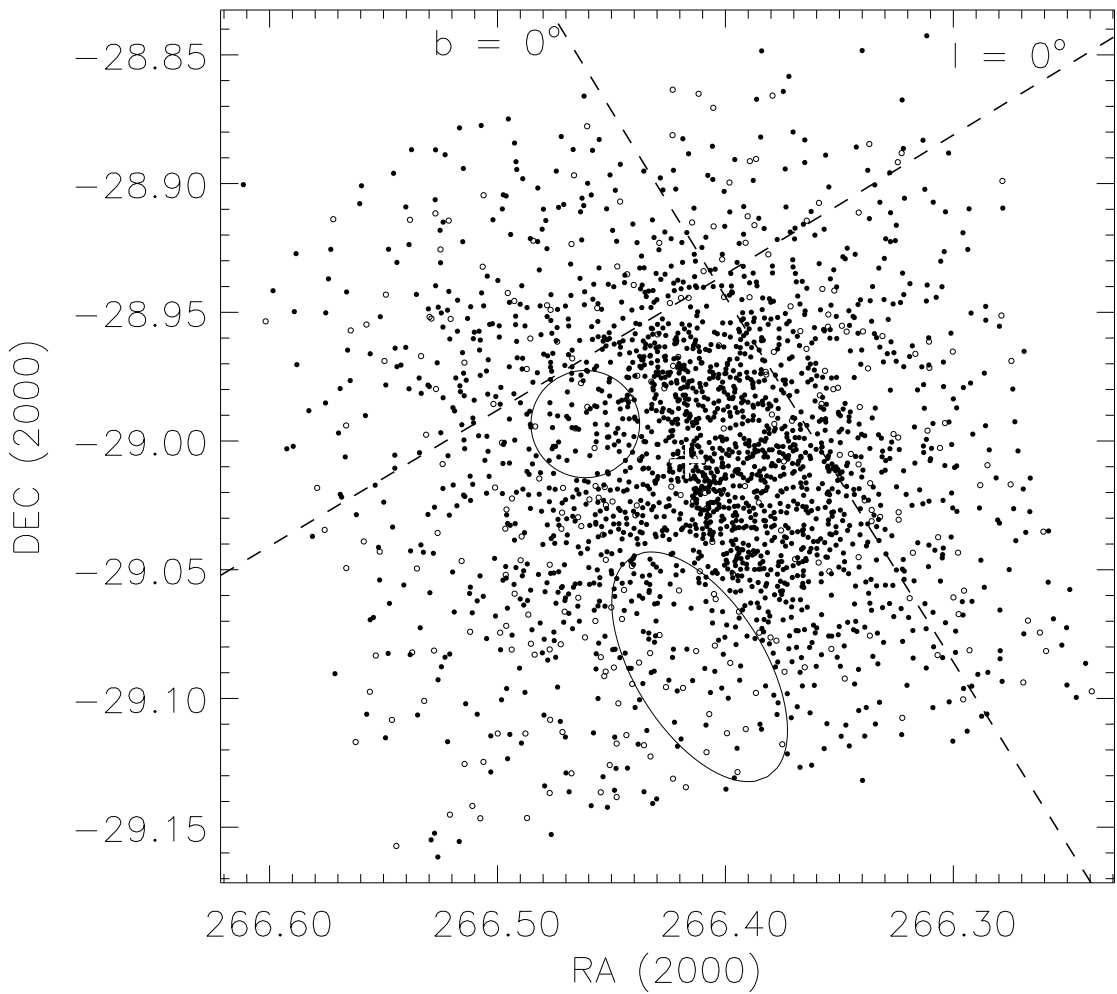,width=0.85\linewidth}}
\caption{The spatial distribution of sources detected in the full 17\arcmin\
by 17\arcmin\ field. Foreground sources (detected below 1.5~keV) are
indicated with open circles, while Galactic center sources are indicated
with filled circles. No attempt has been made to correct the distribution
for the decline in sensitivity at large offsets from the aim point 
(approximately at the center of the image). Galactic longitude and 
latitude of 0\degree\ are plotted with the dashed lines, to indicate the 
orientation of the Galactic plane. \sgrastar\ is indicated with a white 
cross. Ellipses have been drawn to indicate the approximate sizes and 
locations of the two molecular clouds seen within a few arcminutes of 
the Galactic center (see text): 
M$-0.02-0.07$ ({\it upper ellipse}) and M$-0.13-0.08$ ({\it lower ellipse}).}
\label{fig:spatial}
\end{figure*}
that position. We find that this 
strikes a good balance between preventing spurious associations between 
unrelated sources given the uncertainties on the source positions (see the
next paragraph), and our ability to compute the photometry separately for 
two nearby sources (see Section~\ref{sec:phot}). Given the increase in 
density of sources that we find in this field toward \sgrastar\ 
(see Section \ref{sec:spat}) 
and the size of the PSF as a function of field offset angle, there is a 
1\% chance that a second source will lie within one-half the radius of 
the PSF for any given source near \sgrastar, declining to 0.1\% at large 
field offset angles.
Finally, we manually removed a few dozen sources that were obviously part 
of extended, filamentary X-ray features.

With the above method, we found a total of 2357 X-ray point sources. 
Of these, 1792 are detected in the full band, 281 in the soft band 
(124 are exclusively in the soft band), and 1832 in the hard band (441 
exclusively in the hard band). Only 19 sources are detected in all of the 
soft, hard, and full bands. 
Since the absorption column toward the Galactic center is very high
\citep[$6 \times 10^{22}$ cm$^{-2}$ of H; see][]{bag02}, we expect
that very few sources at the Galactic center will be detected below 
1.5 keV. For instance, using the Portable Multi-Mission Simulator 
(\program{PIMMS}), we estimate that a $10^{35}$ \ergsec\ source with a 
6~keV thermal plasma spectrum (e.g. a bright binary system containing
two Wolf-Rayet stars in Table~\ref{tab:ps}) that is absorbed by a column of 
$6 \times 10^{22}$ cm$^{-2}$ and scattered by an equal column of dust
will produce only 3 counts between 0.5-1.5~keV in a 600 ks ACIS-I 
observation. Therefore, even the brightest soft sources are undetectable 
below 1.5 keV if they lie at the Galactic center. For 
the remainder of the paper, we refer to the 281 sources detected
in the soft band as foreground sources, and the rest as 
sources at or beyond the Galactic center.  
We have listed the locations of the 2357 point sources in the electronic 
version of Table~\ref{tab:cat}; the print version
lists the brightest 25 sources in order to provide a sample of the contents
of the full table.

The accuracy of the positions of individual
sources varies significantly over the field, because of variations in the 
size and shape of the PSF. We have cross-correlated the foreground X-ray
sources with stars from the USNO catalog, and find that 148 X-ray sources
have optical counterparts, on order 20 of which could be spurious. 
From the distribution of offsets between the optical and X-ray matches, 
we estimate that the uncertainties on the positions in Table~\ref{tab:cat}
are as small as 0\farcs3 within 1\arcmin\ of the aim point, 
about 0\farcs5 at 4\arcmin, and as large as 2\arcsec\ -- 5\arcsec\ at 
8\arcmin\ -- 12\arcmin\ from the aim point. This is consistent with 
the results of \citet{bra01} 
\begin{figure*}[t]
\centerline{\epsfig{file=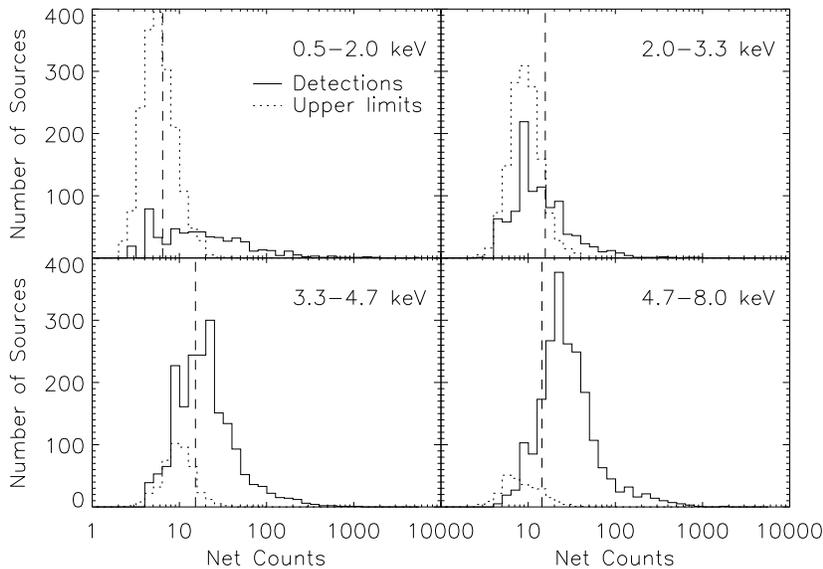,width=0.625\linewidth}}
\caption{Histograms of the number of sources as a function of 
the net counts in each energy band when sources are detected at the 
90\% confidence level ({\it solid line}), and as a function of the
upper limits on the net counts when they are not ({\it dotted line}). 
The vertical {\it dashed line} represents the median number of background 
counts in the source regions. The histograms of detections and upper limits
overlap because of the varying background and PSF size over the image,
both of which determine our sensitivity to faint sources.}
\label{fig:count}
\end{figure*}
for the \chandra\ Deep Field North, and 
\citet{fei02b} for the Orion Nebula. The spatial distribution of sources is 
indicated in Figure~\ref{fig:spatial}, where filled circles denote 
Galactic center sources and open circles denote foreground sources.

\subsection{Photometry\label{sec:phot}}

We used the \program{acis\_extract} routine \citep{bro02} from the Tools for 
X-ray Analysis (TARA\footnote{\html{www.astro.psu.edu/xray/docs/TARA/}}) 
to compute the photometric properties of each source.
We extracted event lists for each source for each observation, using a 
polygonal region generally chosen to match the contour of 90\% encircled energy
from the PSF. We used a PSF at the fiducial energy of 1.5~keV
for foreground sources, while we used a larger extraction area corresponding 
to an energy of 4.5 keV for Galactic center sources.
If two sources were separated by less than twice the 
radius of the 90\% contour, we extracted source counts from smaller 
regions that did not overlap each other. The smallest extraction region 
we used matched the 70\% encircled energy contour; we note that the 
photometry for sources extracted using this contour could be 
inaccurate due to source confusion.
Both the PSF fraction and PSF energy are listed in Table~\ref{tab:cat}.

For each source, a background event list was extracted from a circular region 
centered on the point source, excluding from the event list ({\it i}) 
counts in circles circumscribing 
the 95\% contour of the PSF around any point sources and ({\it ii}) bright, 
filamentary structures. The size of each background
region was chosen such that it contained approximately 1200 total events 
for the 12 observations. We also computed the
effective area function (ARF) and exposure time at the position of each 
source for each observation. We corrected the ARF to account for the fraction 
of the PSF enclosed by the extraction region, and for the hydrocarbon 
build-up on the 
detectors\footnote{\html{cxc.harvard.edu/cal/Acis/Cal\_prods/qeDeg/}}.

The source and background event lists were used to compute photometry for 
each source in four energy bands: 0.5--2.0~keV, 2.0--3.3~keV, 
3.3--4.7~keV, and 4.7--8.0~keV. 
The first band was chosen with an upper limit 
of 2~keV, because the shape of the ARF presents a natural break due to the
telescope's Ir edge, and because below this value most sources at the 
Galactic 
center are dim due to absorption. To define the three higher energy bands, 
we summed all of the counts in the Sgr A* field above 2~keV, and divided 
them into three energy bands with equal numbers of counts. We note that
this results in unconventional boundaries for our energy bands. In any case,
the extremely high absorption toward the
Galactic center would make it difficult to compare our results to those in 
other fields, such as the extra-galactic deep fields \citep{bra01, ros02} 
or globular clusters \citep{pool02}. The total counts, 
estimated background, and mean value of the ARF in each energy band
are listed for the entire sample in the machine-readable version of 
Table~\ref{tab:cat}.

The net counts in each energy band were computed from the 
total counts in the source region less the estimated background.
The uncertainties on the net counts were computed by summing the squares of 
the 1-$\sigma$ upper limits \citep[see Equation 9 in][]{geh86} from both the 
source and background counts. We also computed 90\% confidence intervals 
through a Bayesian analysis, with the simplifying assumption that the 
uncertainty on the background was negligible \citep{kbn91}. If the 90\%
confidence interval on the net counts was consistent with 0, we used
the 90\% upper limit as the uncertainty, 
\begin{inlinefigure}
\centerline{\epsfig{file=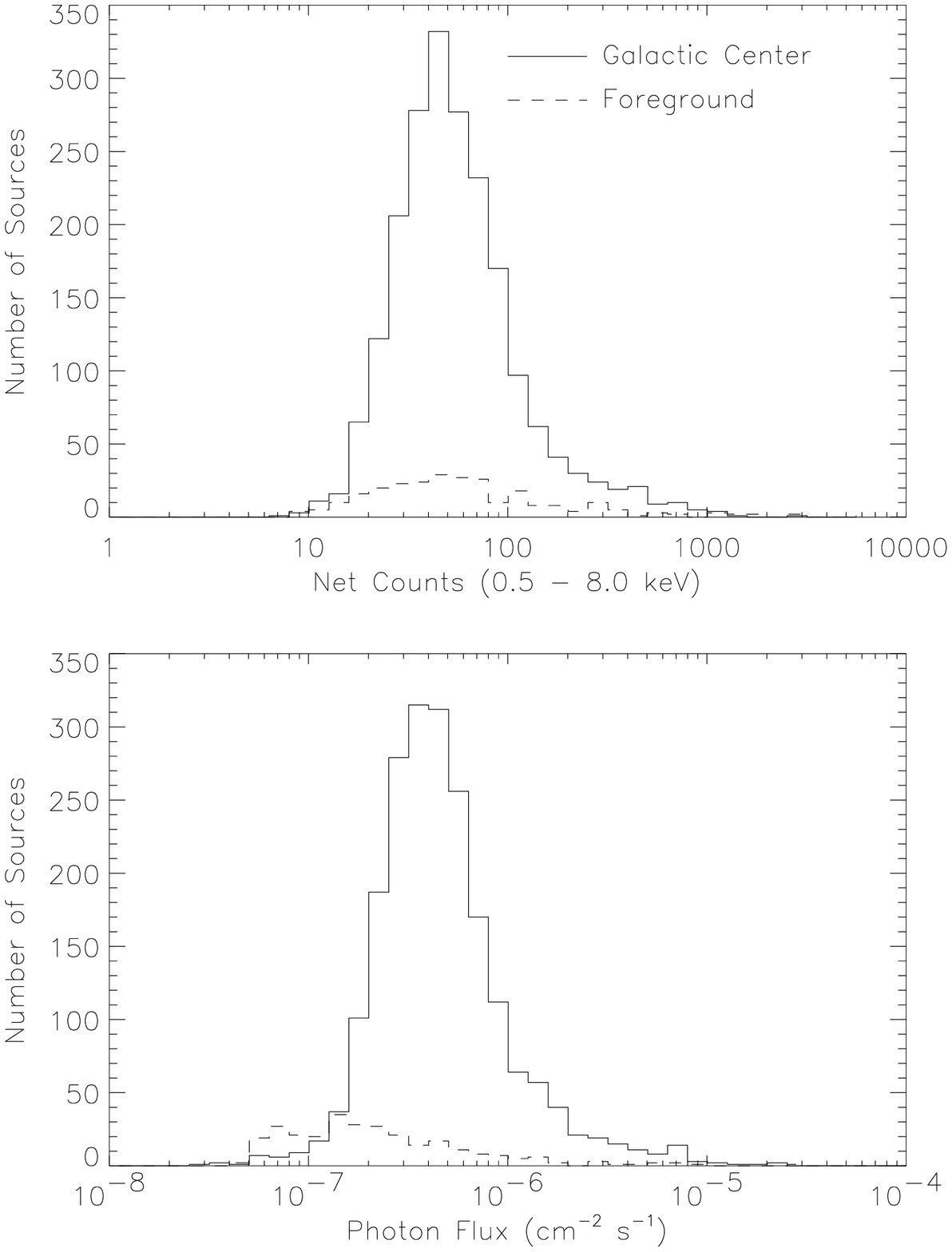,width=0.9\linewidth}}
\caption{{\it Top panel:} Histogram of the number of sources as a function
of the net counts detected in the full band (0.5--8.0~keV). {\it Bottom panel:}
Histogram of the number of sources as a function of photon flux in the 
full band. Galactic center sources are indicated with the {\it solid lines},
for which 1~\phcms~$= 8\times10^{-9}$~\ergcms\ (2.0--8~keV).
Foreground sources (detected below 1.5~keV) are indicated with the 
{\it dashed lines}, for which 1~\phcms~$= 3\times10^{-9}$ \ergcms\
(0.5--2.0~keV). The luminosities of the Galactic center sources
range between $10^{30}$ and $10^{33}$ \ergsec\ for a distance of 8~kpc
\citep{mcn00}.}
\label{fig:lumdist}
\end{inlinefigure}

\noindent
although we retained the most
likely flux even if it was negative. Histograms of the number of sources
as a function of the net counts in each energy band are displayed in
Figure~\ref{fig:count} ({\it solid lines}). Those sources with only 
upper limits are represented with the {\it dotted lines}, and the median
numbers of background counts in the source regions are indicated with the 
vertical {\it dashed lines}. Note that the photometry for sources with
a small number of net counts in a given energy band (typically less than 
5 counts) may be unreliable, as there are comparable systematic 
uncertainties in the background estimates due to the spatially varying 
diffuse emission. 
The net counts in the full 0.5--8.0~keV energy band are listed for the
entire sample in the electronic version of Table~\ref{tab:cat}, and
for the brightest 25 sources in the print version. A histogram of the 
number of foreground ({\it dashed line}) and Galactic center 
({\it solid line}) sources as a function of the net counts is displayed in 
the top panel of Figure~\ref{fig:lumdist}.

We computed approximate photon fluxes (in units of \phcms) for each source
by dividing the net counts in each sub-band by the total live time
(units of s) and the mean value of the ARF in that energy range 
(units of cm$^2$; note that this value incorporates variations in exposure due 
to chip gaps and dead columns). The photon fluxes in each band are listed in 
Table~\ref{tab:cat}, along with uncertainties or upper limits. The photon 
fluxes in the 2.0--8.0~keV energy band used throughout the paper are the 
sums of those in the sub-bands, using negative values when they occur 
(not the upper limits). Since the energy bands sampled
the ARF for the ACIS-I detector well, the approximate photon 
fluxes that we computed differed from those derived from later spectral fits 
using \program{XSPEC} (Section~\ref{sec:spec}) by no more than the 
uncertainty expected from Poisson counting noise. 

A histogram of the number of sources as a function 
of the 0.5--8.0~keV photon flux is presented in the {\it bottom panel} of 
Figure~\ref{fig:lumdist}. Galactic center sources are indicated with the 
{\it solid line}, and foreground sources with the {\it dashed line}.
Sources are detected with 
photon fluxes as low as $5\times10^{-8}$~\phcms. The largest number of 
Galactic center sources is detected near $4\times10^{-7}$~\phcms\
(2.0--8.0~keV), and the largest number of foreground sources is found near 
$1\times10^{-7}$~\phcms\ (0.5--2.0~keV). 
Since we become less sensitive to detecting sources at 
positions far from the aim point (see Section~\ref{sec:area}), the 
smaller number of sources at low fluxes probably occurs because of 
incompleteness.

We used the spectral models from Section~\ref{sec:spec} to compute an 
average conversion factor
between photon and energy flux. For Galactic center sources (not detected 
below 1.5~keV), we find 
that for the typical $\Gamma = 0.5$ power-law spectrum absorbed by a
column equivalent to $6\times10^{22}$ cm$^{-2}$ of H, an absorbed flux 
of 1~\phcms~$= 8\times10^{-9}$~\ergcms\ (2.0--8.0~keV). 
The unabsorbed flux 
is approximately 50\% larger. For sources detected below 1.5~keV, we
find that 1~\phcms~$= 2\times10^{-9}$ \ergcms\ between 0.5--2.0~keV. 
The absorption for these sources is relatively small ($< 10^{22}$ cm$^{-2}$).

\subsection{Solid Angle of the Survey \label{sec:area}}

In order to make quantitative statements about the spatial and luminosity
distributions of the sources in this sample, it was necessary to understand
the limiting flux at which we could reliably measure the flux from a source 
as a function of position on the sky. The signal-to-noise ratio with 
which we measure the flux from a source is 
$n_{\sigma} = N/[(N + B) + \sigma_B^2]^{1/2}$, 
where $N$ is the net number of counts from a source, and $\sigma_B$ is
the uncertainty on the background $B$ in the source region.
This definition is simply the flux divided by its uncertainty in 
Table~\ref{tab:cat}, although here we make the simplifying 
assumption of $\sqrt{N}$ uncertainties. The net counts are 
related to the flux from the source $S$ by $N = SAT$, where $A$ is the 
effective area of the detector, and $T$ is the exposure time. For simplicity,
the background 
can be written as the product of the background per pixel $b$ and the area of 
the PSF $a$, $B = ba$. To obtain a conservative estimate of our flux
limit, we will take $\sigma_B^2 = B$, although in practice we determine
the background over an area that is $5 - 30$ times larger than the source
extraction region, which lowers $\sigma_B^2$ significantly. 
 The signal-to-noise of the source then can be written as
\begin{equation}
n_{\sigma} = {{SAT} \over {(SAT + 2ba)}^{1/2}}.
\label{eq:ston}
\end{equation}
If we observe down to a well-defined signal-to-noise, we can then invert 
Equation~\ref{eq:ston} to derive position-dependent flux limits for our 
image
\begin{equation}
S = {{n_{\sigma}^2} \over {2}} {{1} \over {AT}} 
    \left( 1 + \left[1 + {{8ba}\over{n_{\sigma}^2}} \right]^{1/2} \right)
\label{eq:fluxlim}
\end{equation}
\begin{figure*}[t]
\centerline{\epsfig{file=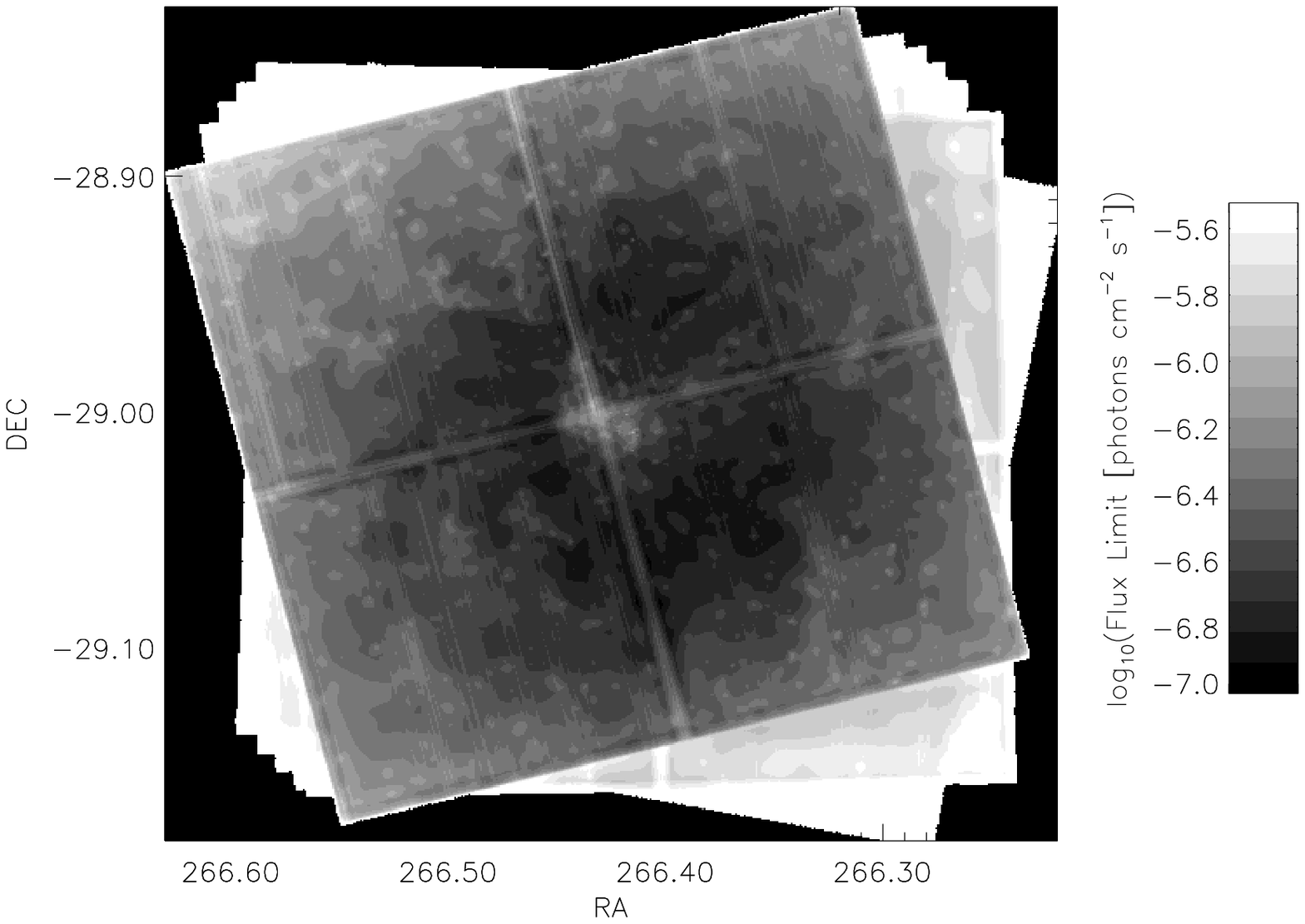,width=0.8\linewidth}}
\caption{Map of the limiting flux in the full band for our observations,
computed according to Equation~2 assuming a signal-to-noise of 3.0 on 
a flux measurement.
The key to the grey scale is indicated to the 
right of the image; black indicates the best sensitivity.
 Various effects limiting our sensitivity are 
evident, including the presence of chip gaps and bright diffuse emission, and
the increase in the PSF size as a function of offset from the aim point.}
\label{fig:fluxlim}
\end{figure*}
\begin{inlinefigure}
\centerline{\epsfig{file=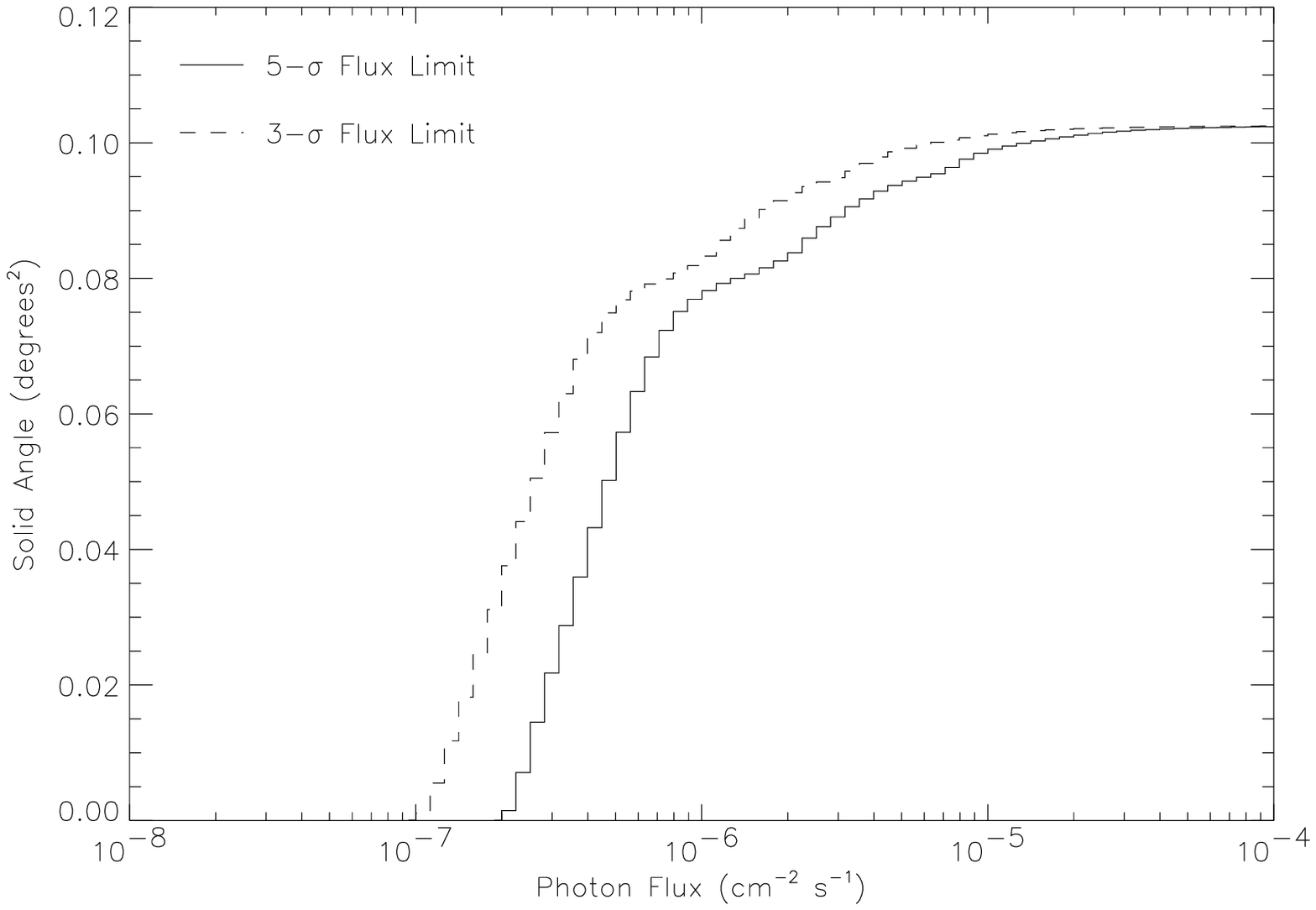,width=\linewidth}}
\caption{Plot of the solid angle observed at each limiting flux,
for the cases when the flux can be measured with a signal-to-noise of 
3 (dashed line) and 5 (solid).
}
\label{fig:area}
\end{inlinefigure}

\noindent
\citep[compare][]{man03}.
Figure~\ref{fig:fluxlim} illustrates a map of the limiting flux available 
at each position in our image for $n_{\sigma} = 3$, as defined by 
Equation~\ref{eq:fluxlim}. 

The map takes into account the varying background 
level, exposure, effective detector area, and PSF area over the field. 
The background level was estimated by removing circular regions
containing the point sources, filling the resulting holes in the image
with a Poisson distribution of counts that matches a surrounding annulus
using \program{dmfilth}, 
and smoothing the final image using the routine \program{csmooth}. The diffuse 
features in the background increase the flux limit, which is particularly
evident around the Sgr A complex just below and to the right of the center
of the image.
The exposure and effective area were estimated using standard 
CIAO tools. These produce the bright cross due to the gaps between the CCDs
and the vertical stripes due to bad columns excluded in the analysis. The PSF 
area was determined from the 90\% encircled energy contours used to 
extract the point sources. The increasing size of the PSF with field offset
angle causes the steady increase in limiting flux at the edges of the image. 

Figure~\ref{fig:area} displays 
the solid angle observed as a function of limiting flux for 
$n_{\sigma} = 3$ and
$n_{\sigma} = 5$, which we computed by summing the pixels in the flux
map image (Figure~\ref{fig:fluxlim}). The solid angle of our
survey at low fluxes is strongly dependent on our assumed signal-to-noise.
We believe that we have detected all sources with $n_{\sigma} > 3$, 
based
on the peak in the number of sources detected as a function of flux
(Figure~\ref{fig:lumdist}).
On the other hand,  $n_{\sigma} > 5$ is the limit at which we know the 
flux from a source with enough accuracy to determine the slope of the 
number
count distribution (Murdoch, Crawford, \& Jauncey 1973)\nocite{mcj73}. 

\section{Results\label{sec:res}}

\subsection{Spatial Distribution\label{sec:spat}}

\begin{figure*}[t]
\centerline{\epsfig{file=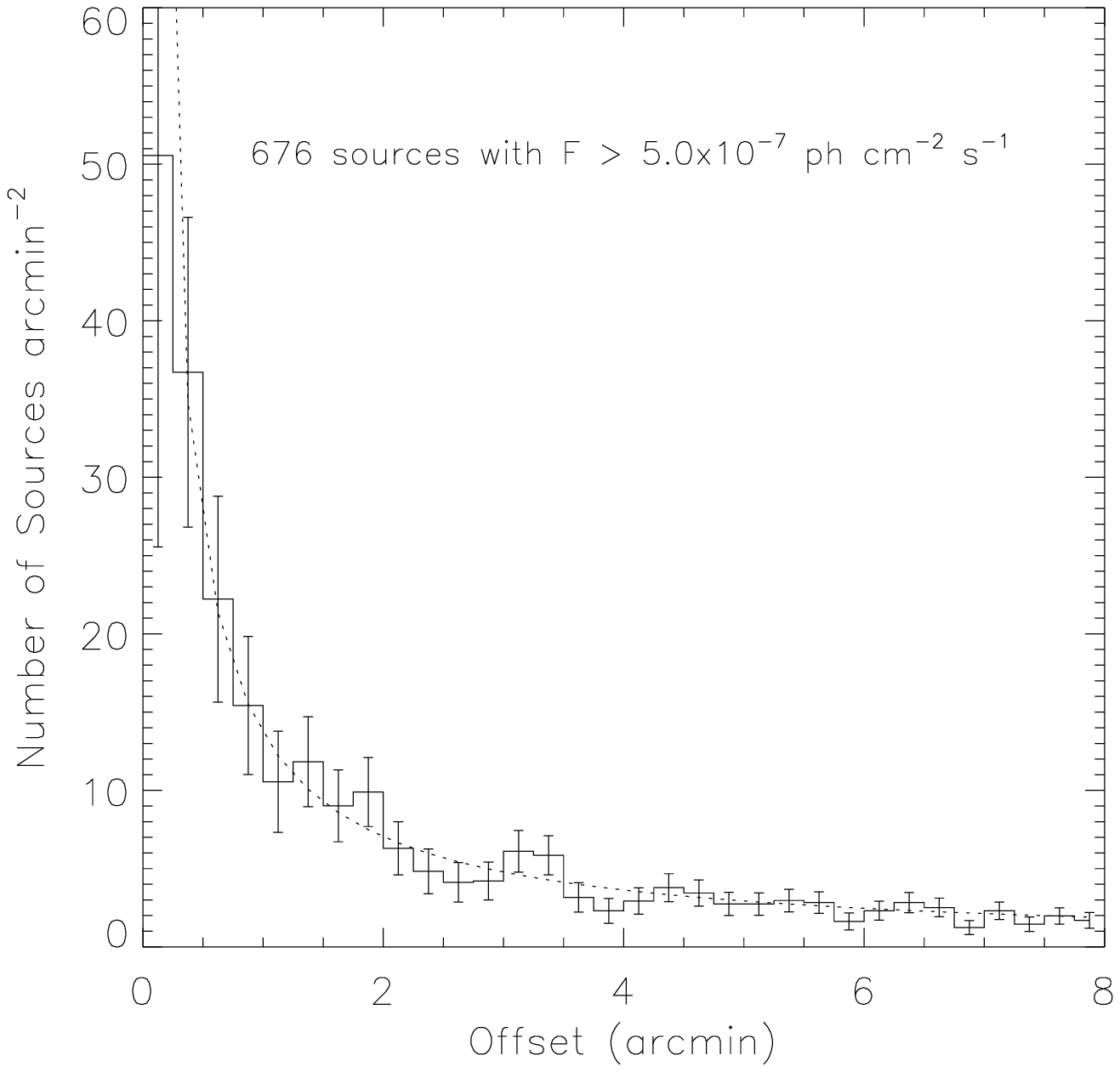,width=0.75\linewidth}}
\caption{Surface density of Galactic center point sources as a function
of offset angle from Sgr A*. The number of 
sources in each annulus was divided by the solid angle over which a source 
could be detected above $5\times10^{-7}$ \phcms\
with a signal-to-noise of 3 in that annulus. The dotted line indicates a 
fit of $\theta^{-\beta}$
decrease in surface density with increasing offset $\theta$, where 
$\beta = 1.0\pm0.1$.
}
\label{fig:radialprofile}
\end{figure*}
Figure~\ref{fig:spatial} illustrates the spatial distribution of sources
detected in the field, without correcting for the decreasing sensitivity 
of the instrument at larger field offsets from the aim point. An east-west 
asymmetry is evident about the Galactic center. The deficit of sources 
to the east of \sgrastar\ corresponds to the locations of two molecular 
clouds, M$-0.02-0.07$ and M$-0.13-0.08$, which are indicated 
schematically by the two ovals in Figure~\ref{fig:spatial} 
(G\"{u}sten, Walmsley, \& Pauls 1981; Mezger \etal\ 1996)\nocite{gwp81,mdz96}. 
The clouds are thought to 
lie in or in front of the Galactic center (Zylka, Mezger, \& Wink 
1990)\nocite{zmw90}, and therefore 
probably obscure X-ray sources that lie beyond them.
They have mean column densities of 
$N_{\rm H} \approx 3\times10^{23}$~cm$^{-2}$, which is a factor of five
higher than the mean Galactic value \citep{zmw90}. This would reduce 
the observed flux in the \chandra\ bandpass by 30\%, which would decrease
the number of sources observed by 65\% given the luminosity distribution 
of Galactic center sources (see Section~\ref{sec:flux}). Fewer sources are also
detected near Sgr A East, probably because the strong diffuse emission 
masks the emission from point sources. 

At about 8\arcmin\ from the aim point, we estimate that we can detect all 
sources with a photon flux greater than $5\times10^{-7}$ \phcms\ with a 
signal-to-noise of at least $n_{\sigma} = 3$ in the 2.0--8.0~keV band (compare
Figures~\ref{fig:lumdist}, \ref{fig:fluxlim}, and \ref{fig:area}). 
About 40\% of the
Galactic center sources detected within 8\arcmin\ of \sgrastar\ have photon 
fluxes greater than this value. 
In Figure~\ref{fig:radialprofile} we plot the 
number of Galactic center sources above this flux limit per unit solid angle 
as a function of angular separation from Sgr A*. We have fit this 
distribution with a power law of the form
\begin{equation}
\Sigma(\theta) = (14\pm3) \theta^{-1.0\pm0.1} {\rm sources~arcmin}^{-2},
\label{eq:rad}
\end{equation}
where $\theta$ is the angular separation in arcminutes. 
Both the normalization and 
the power-law slope were allowed to vary. The resulting fit is acceptable, 
with a $\chi^2$ of 26 for 30 degrees of freedom. If we assume that these 
sources are distributed with spherical symmetry about the Galactic center, 
the implied spatial density falls off with radius as $R^{-2}$. 

The background in the soft band is much lower, so foreground sources can
be detected reliably down to a limit of $1.6\times10^{-7}$ \phcms\ in
the 0.5--2.0~keV band with a signal-to-noise of $n_{\sigma} = 3$.
There are 120 soft sources detected above this flux limit in the inner 
8\arcmin\ of the field.
We have compared the radial distribution of foreground sources to 
a uniform distribution using a KS-test, and find that there is only a 
55\% chance that the two are different. Assuming the distribution is 
uniform, the mean surface density is $0.5$ sources arcmin$^{-2}$, with a 
standard deviation of 0.2 sources arcmin$^{-2}$. 
The foreground sources probably all reside in the Galactic disk, which should 
exhibit no significant radial density gradient in this 17\arcmin\ field.
However, some inhomogeneities in the surface density
should be expected from spatial variations in the absorption column 
\begin{figure*}[t]
\centerline{\epsfig{file=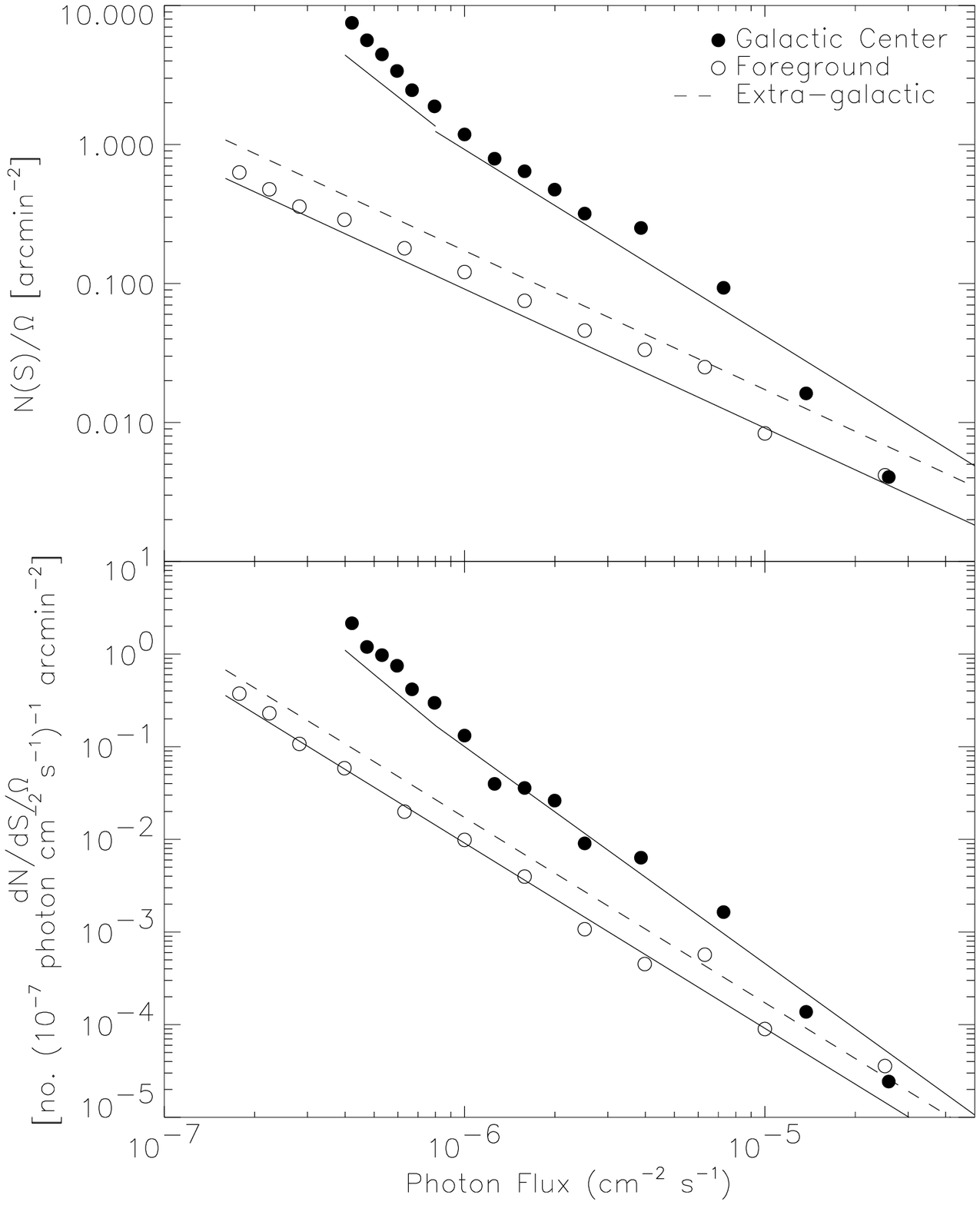,width=0.5\linewidth}}
\caption{Cumulative $\log(N)-\log(S)$ distribution ({\it top panel})
and differential number counts ({\it bottom panel}) of sources at the 
Galactic center (filled circles) and in the foreground (open circles). The
distributions have been  
normalized according to 
({\it i}) the solid angle available in each flux range as
listed in Table~4 (compare Figure~4), and 
({\it ii}) the reduced surface density 
that would be expected for Galactic center sources over the entire 
9\arcmin\ survey, given
that the density of sources falls off as $1/\theta$. The best-fit models 
determined by a maximum-likelihood method are over-plotted with solid
lines. At low fluxes, the model prediction for the number of Galactic center 
sources appears to be low; this is because the distribution is extremely
steep, and the uncertainties on the flux measurements will tend to 
preferentially shift sources from lower flux bins to higher ones (Eddington 
bias). The expected extra-galactic contribution from \citet{bra01} 
is indicated with the dashed line \citep[see also][]{ros02}. We note that 
1~\phcms~$= 8\times10^{-9}$~\ergcms\ (2.0--8~keV) for Galactic center sources,
and 1~\phcms~$= 3\times10^{-9}$ \ergcms\ (0.5--2.0~keV) for foreground
sources. }
\label{fig:norm_dnds}
\end{figure*}
throughout this field.

\subsection{Flux Distribution\label{sec:flux}}

In Figure~\ref{fig:norm_dnds}, we plot both the cumulative
and the differential
number counts as a function of flux for sources at the 
Galactic center (filled
circles) and in the foreground (open circles), normalized to the solid 
angle of the survey $\Omega$ 
in units of arcmin$^{-2}$. We have been conservative in our source selection 
to avoid incompleteness in our sample caused by the varying sensitivity 
over our image (Figure~\ref{fig:fluxlim}). Table~\ref{tab:surveys} lists
the criteria used to select sources for this distribution. First, sources 
in a given flux range are required to lie within a maximum field offset angle. 
This criterion accounts for the increasing PSF size with field offset, which 
causes our sensitivity to decrease.
Second, the $n_{\sigma} = 5$ flux limit derived from Equation~\ref{eq:fluxlim}
at the position of each source is required to be smaller than the value in
Table~\ref{tab:surveys} for that field offset angle. This excludes 
regions of high background, detector chip gaps, and bad CCD columns. 
The solid angles over which sources from each flux limit are acceptable are 
listed in Table~\ref{tab:surveys}, as well as the number of sources accepted 
in each flux range. Note that only about one-third of both the foreground
and Galactic center sources detected in our image satisfy the above 
selection criteria, since the high background in the image adds significant
uncertainty to our flux measurements. The $dN/dS$ 
distribution is constructed from the number of sources in a flux interval
$dN$, divided by the size of the flux interval $dS$, divided by the solid
angle of the survey at that flux. The flux interval is computed in units of 
$10^{-7}$ \phcms. Finally, since the surface density of Galactic center
sources decreases with increasing field offset, we normalized the numbers of 
hard sources in each flux range to the mean surface density in the 
largest 9\arcmin\ survey area. The cumulative distribution $N(S)$ is 
simply the integral of the differential distribution.

Using the un-binned flux values, we modeled the $\log N - \log S$ 
distributions using the maximum likelihood technique described in 
\citet{mcj73}. 
The Galactic center sources were not consistent with a
single power-law
distribution of the form
\begin{equation}
N(S) = N_o \left( {{S} \over {4\times 10^{-7}}} \right)^{-\alpha}.
\label{eq:pow}
\end{equation}
Therefore, guided roughly by the differential counts in 
Figure~\ref{fig:norm_dnds}, we divided the distribution into two flux ranges, 
less than and greater than $8\times10^{-7}$ \phcms. These regions were 
consistent with power laws of the forms 
\begin{equation}
 N(S) = \\ \left\{ \begin{array}{ll}
 (4.4\pm0.2) \left( {{S} \over {4\times 10^{-7}}} \right)^{-1.7\pm 0.2} &
S < 8.0 \times 10^{-7} \\
(1.11\pm0.07) \left( {{S} \over {8\times 10^{-7}}} \right)^{-1.34\pm 0.08} &
S > 8.0 \times 10^{-7}
\end{array} \right.
\label{eq:mod}
\end{equation}
where $S$ is in units of \phcms, and the normalization is in units of 
sources arcmin$^{-2}$. 
The normalizations of each model distribution are set to the mean value that
would be expected over the full 9\arcmin\ survey area, given that the observed
stellar density falls off approximately as $\theta^{-1}$. 
We take Equation~\ref{eq:mod} as a rough approximation of 
the $\log(N) - \log(S)$ distribution of the Galactic center sources.

Since the surface density of hard sources is strongly peaked at the Galactic 
center, it is reasonable to assume that most of them are within the nuclear
bulge and hence at the same distance, $8.0\pm0.3$ kpc 
\citep[][see also Section~3.1]{mdz96}.
As a result, 
if the photon fluxes can be converted to unabsorbed energy fluxes, this 
cumulative number--flux distribution would represent the intrinsic
luminosity distribution of the sources. Unfortunately, the absorption varies 
significantly from source to source, so significant uncertainties would 
be introduced in converting this to a luminosity distribution. Moreover, 
the physical meaning of the slopes of the distributions is at the moment
unclear, since we have not yet identified the nature of the sources. 
Therefore, we have made no attempt to match the 
normalizations of these two model 
distributions exactly, nor did we attempt to derive  more rigorously the 
position of the break in the distribution. 

The flux distribution of the foreground sources in
Figure~\ref{fig:norm_dnds} (open circles)
was similarly modeled. Only the
soft band (0.5--2.0~keV) was used in computing the flux, since this 
minimizes the background. The 
power law determined from the maximum likelihood technique was 
\begin{equation}
N(S) = 
(0.57\pm0.08) \left( {{S} \over {1.6\times 10^{-7}}} \right)^{-1.00\pm 0.09},
\label{eq:modsoft}
\end{equation}
where $S$ is again in \phcms, and the normalization is sources arcmin$^{-2}$.
This distribution is much flatter than that at the Galactic center. 

\subsection{Confusion Limit\label{sec:conf}}

Despite the large number of sources in this image, we are still far 
from being confusion-limited. \citet{hogg01} demonstrated that for 
steep ($\alpha = 1.5$) number count distribution, fewer than 10\% 
of sources will be affected by confusion if there are fewer than one source
per 50 ``beams''. Here, a beam is defined to have an area 
$\pi(\theta_{\rm FWHM}/2.35)^2$, where $\theta_{\rm FWHM}$ is the 
full-width half-maximum of the PSF. In the inner 1\arcmin\ of the 
image in Figure~\ref{fig:rawimage}, 
there are 13 sources arcmin$^{-2}$ with fluxes higher than $5\times10^{-7}$
\phcms\ (Equation~\ref{eq:rad}). Since $\theta_{\rm FWHM}$ for \chandra\
is only 0\farcs5 near the aim point, there is approximately once source 
per 2000 beams in the densest part of our image. The confusion limit will 
not be reached until the surface density increases by a factor of 40. From
Equation~\ref{eq:mod}, we would have to observe to a completeness limit
a factor of 9 lower in order for confusion to be important. 
On the other hand, at an offset of 8\arcmin from the aim point, the
spatial density of sources is 1.7 per arcmin (Equation~\ref{eq:rad}), and
\begin{inlinefigure}
\centerline{\epsfig{file=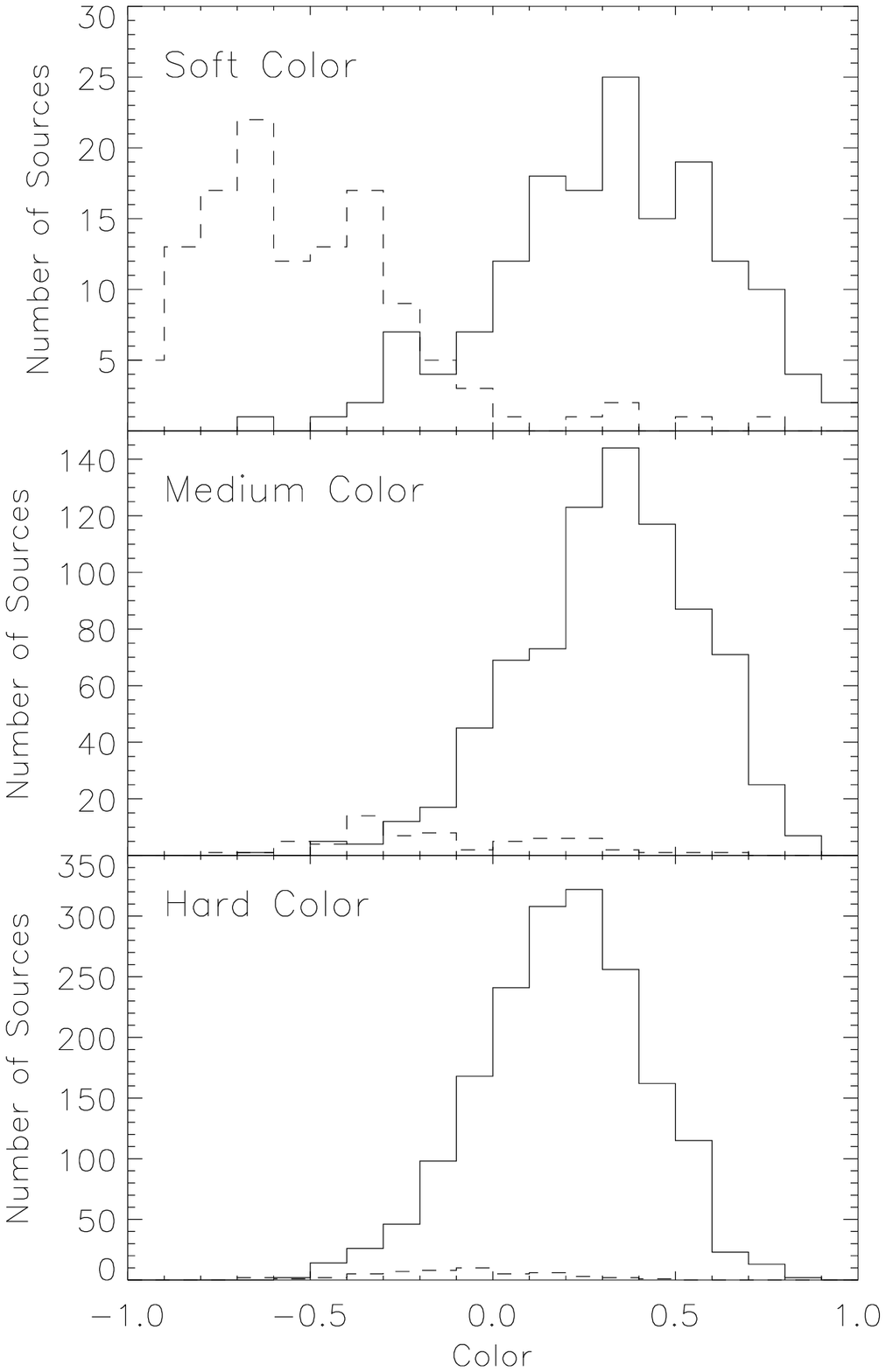,width=0.9\linewidth}}
\caption{
Histogram of hardness ratios from sources that are detected with greater than
90\% confidence in each relevant energy band. 
The hardness ratios are defined according to $(h-s)/(h+s)$, where $s$ are the 
counts in a low energy band, and $h$ are the counts in a high energy band.
The soft color is  
defined as the fractional difference between counts with energies
between 2.0--3.3~keV and 0.5--2.0~keV, the medium color using counts 
between 3.3--4.7~keV and 2.0--3.3~keV, and the hard color using counts 
between 4.7--8.0~keV and 3.3--4.7~keV. Foreground 
sources are indicated with the dashed histogram, while Galactic center 
sources are indicated with the solid histogram. 
}
\label{fig:hr}
\end{inlinefigure}

\noindent
$\theta_{\rm FWHM} = 5$\arcsec, which implies that there is one source
per 200 beams. Thus, the outer edges of the image will become confusion 
limited if our completeness limit reaches a factor of 2 lower.
Therefore, both the photometry and the positions of the X-ray sources
in this sample are reliable over the entire image.

\subsection{Hardness Ratios\label{sec:hr}} 

We used the counts in each energy band to compute three hardness ratios,
which we used to characterize the absorption column toward each 
source and the steepness of the high-energy portion of each spectrum. 
The ratios are defined 
as the fractional difference between the count rates in two 
energy bands, $(h-s)/(h+s)$, where 
$h$ and $s$ are the numbers of counts in the higher and lower energy bands, 
respectively. The resulting ratio is bounded by $-1$ and $+1$. 
The soft color is defined by the fractional difference between counts 
with energies between 2.0--3.3~keV 
\begin{figure*}[t]
\centerline{\epsfig{file=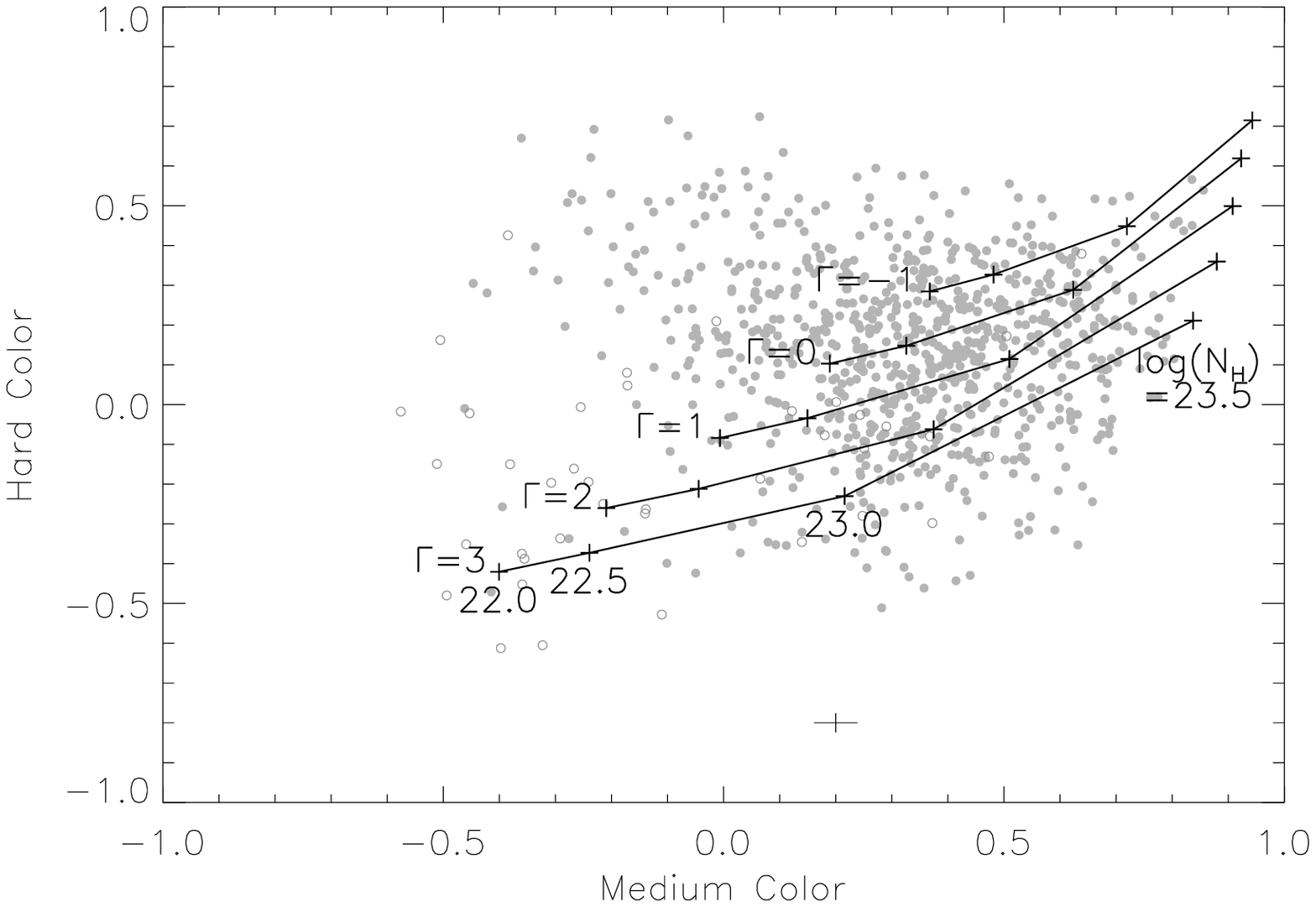,width=0.8\linewidth}}
\caption{Comparison of the observed hard and medium colors to those expected 
from an absorbed power-law spectrum for sources detected with greater
than 90\% confidence in all three energy bands above 2~keV. 
Data from point sources are indicated in 
grey, with open circles denoting foreground sources and filled circles those 
sources at the Galactic center. The crosses connected with solid lines 
indicate the expected colors for absorbed power laws, with the values 
indicated on the plot. 
The median uncertainty for these sources is displayed at the bottom of 
the plot. We note that the sources in the upper-left corner of 
Figure~\ref{fig:cc} all have uncertainties a factor of 2--3 larger.
}
\label{fig:cc}
\end{figure*}
and 0.5--2.0~keV; the medium color using
counts with energies between 3.3--4.7~keV and 2.0--3.3~keV, and the hard 
color using counts between 4.7--8.0~keV and 3.3--4.7~keV. The hardness 
ratios are listed in Table~\ref{tab:cat}, with uncertainties calculated 
according to Equation~1.31 in Lyons (1991; page 26).
Histograms of the hardness ratios for sources that are detected with 
greater than 90\% confidence in each relevant energy band are plotted in 
Figure~\ref{fig:hr}. We indicate separately
sources in the foreground ({\it dashed line}) and at the Galactic center 
({\it solid line}). By design, the foreground sources tend to have lower 
soft colors than the Galactic center sources, indicating more photons are 
received at low energies. Relatively few Galactic
center sources are detected with 90\% confidence in the 0.5--2.0~keV band
by definition, so the soft colors have limited usefulness. The medium and 
hard colors provide more information about Galactic center sources. 

We have calculated the hardness ratios that we would expect
to get from these energy bands for a variety of spectra using \program{PIMMS}. 
Variations in the
soft color are dominated by differences 
in the absorption column toward
the source, although we find that sources with column densities greater than 
$10^{22.5}$ cm$^{-2}$ would have too few counts to be detected reliably below
2~keV. The hard color is determined almost exclusively by the temperature 
or steepness of the spectrum above 4~keV, as long as the absorbing column
is less than about $10^{23.5}$~cm$^{-2}$ of H. The medium color is affected 
by both the absorbing column and the intrinsic spectral shape. 
In order to examine the spectra further, 
in Figure~\ref{fig:cc} we compare the medium and hard colors of the point 
sources with those expected from a set of simulated absorbed power-law 
spectra (of the form $E^{-\Gamma}$). We have indicated Galactic center 
sources by filled circles, and foreground sources with open circles. 
All sources detected with 90\% confidence in the three energy bands above 
2~keV are included in the plot, which amounts to 785 Galactic center sources 
and 39 foreground sources. 

About half of the Galactic center sources cluster in a region consistent 
with absorption columns $\log(N_{\rm H}) > 22.5$, and very flat
spectra with photon indices $\Gamma < 1$ (where negative values indicate 
rising numbers of photons with energy). Such hard spectra are unusual
for X-ray point sources (see Table~\ref{tab:ps}). Simulations
with \program{PIMMS} indicate that the thermal models expected for 
many of the classes of sources in Table~\ref{tab:ps} --- e.g. ionized plasma 
with $kT < 25$~keV, blackbodies with $kT<2$~keV, or Bremsstrahlung emission 
with $kT < 50$~keV --- all produce hard colors less than 0.1. 
We have confirmed that the X-ray spectra of most of the Galactic center sources
are intrinsically hard (as opposed to resulting only from high absorption)
using the spectral fits reported in the next section. 

\subsection{Spectra \label{sec:spec}}

We modeled the spectra of sources
with more than 80 net counts in the full band (approximately $6\times10^{-7}$
\phcms\ between 0.5--8.0~keV) using simple models in 
\program{XSPEC}. The models consisted of a blackbody, bremsstrahlung, or
power-law continuum multiplied by 
\begin{figure*}[t]
\centerline{\epsfig{file=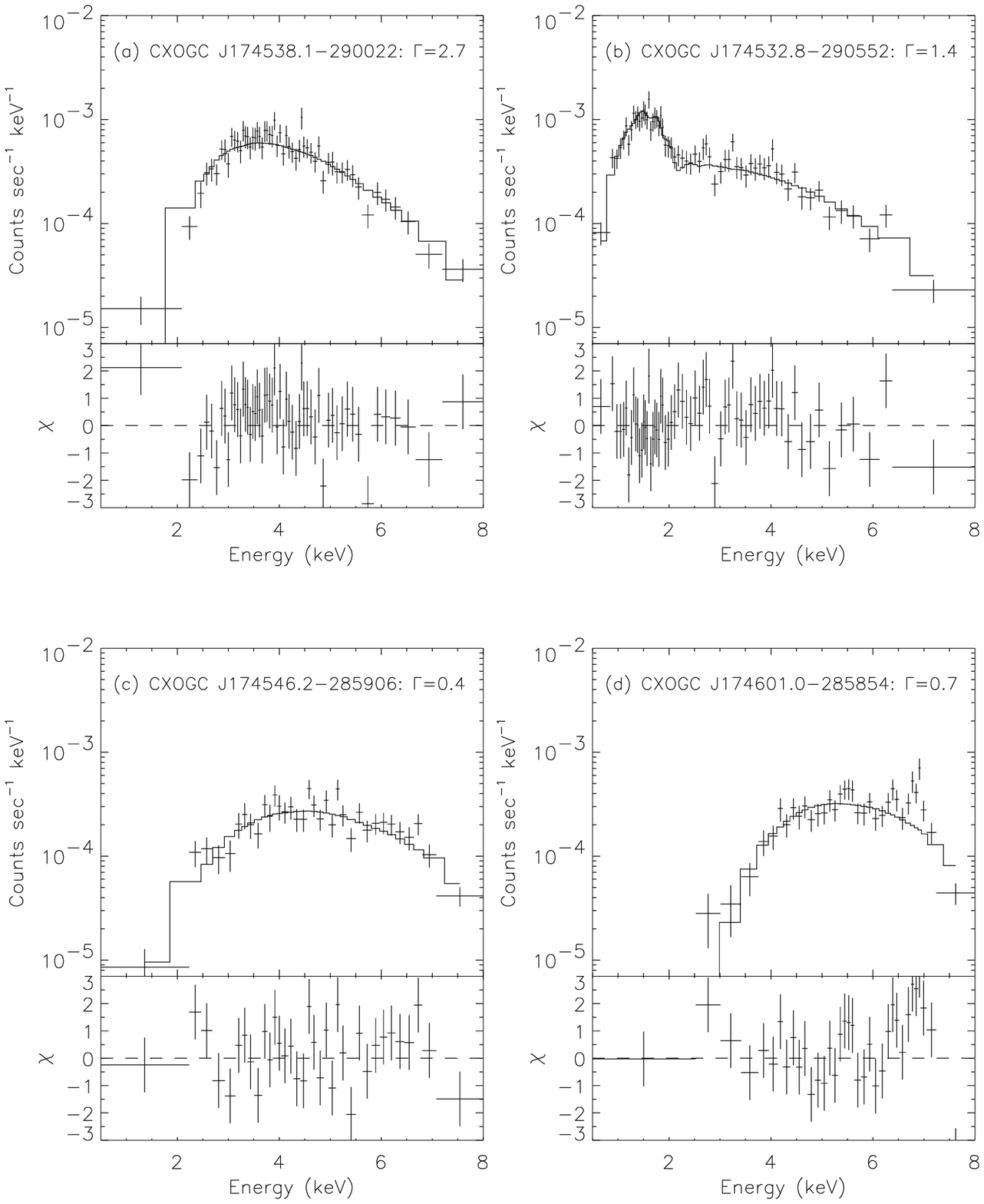,width=0.85\linewidth}}
\caption{Example spectra of four bright sources. The 
{\it top panels} have the spectrum in 
units of detector counts sec$^{-1}$ keV$^{-1}$ as a function of energy
in keV, so that the varying effective area of the detector is convolved
with the spectrum. The solid histograms represent the best-fit power laws,
which are statistically acceptable in all cases except the last. The
{\it bottom panels} show the residuals to the fit, in units of the $\chi$ 
statistic. 
{\it Panel a} illustrates a Galactic center source with a steep $\Gamma = 2.7$ 
power law and hard color $-0.25$. {\it Panel b} illustrates a foreground source with a $\Gamma = 1.4$ 
power law and hard color $-0.26$. {\it Panels c} and {\it d} illustrate the hard, $\Gamma < 1$ 
power law spectra 
measured for many of the Galactic center sources (hard colors 0.16 and 0.56, respectively). In {\it panel d}, there are 
also significant residuals between 6--7~keV due to un-modeled iron emission.}
\label{fig:spec}
\end{figure*}
factors to account for interstellar 
absorption with column density $N_{\rm H}$ and dust scattering with optical
depth $\tau$.  The column depth of dust was set to 
$\tau = 0.485 \cdot N_{\rm H}/(10^{22} {\rm cm}^{-2})$ \citep{bag02}, 
and the halo size to 100 times the PSF size.
We found that most of the sources that we modeled, 436 of 565, could be 
adequately fit (at least a 10\% chance of exceeding the observed 
$\chi^2$ randomly) with an absorbed power law. The power-law models were not 
adequate for brighter sources with strong line emission, and for very soft
sources. The bremsstrahlung and blackbody models succeeded only 
slightly less often, but more than half the time the temperatures obtained
were extremely hot ($kT > 25$~keV for bremsstrahlung or $kT > 3$~keV for
blackbodies).
A full exploration of the spectra of the point sources is beyond the 
scope of the current paper, but these preliminary spectral
fits do confirm the results from our analysis of hardness ratios.

Four example spectra with a range of photon indices are displayed in 
Figure~\ref{fig:spec}. {\it Panel a} displays a highly absorbed source with a 
relatively soft spectrum above 3.5~keV ($\Gamma=2.7$ power law, or a hard 
color of $-0.25$). 
{\it Panel b} exhibits a foreground source, which can be identified by its 
copious 
emission below 2~keV. This spectrum is adequately modeled with a 
power law of photon index $\Gamma=1.4$, and has a hard color of $-0.26$. 
{\it Panels c} and {\it d} illustrate 
spectra from two very hard sources that have best-fit power laws of slope 
$\Gamma=0.4$ (hard color 0.16) and $0.7$ (hard color 0.57), respectively. 
Visual inspection confirms the hardness of the
spectra, because these sources have 
as much or more flux above 5~keV than the other two sources, 
but less flux between 3.3--4.7~keV. The source in 
{\it panel d} also exhibits copious iron emission, so that the power law 
fit to this 
source is statistically unacceptable. Including a line of equivalent width 
0.8~keV at 6.9~keV produces 
an acceptable fit, and increases the photon index to $\Gamma = 1.0\pm 0.8$.
We conclude that hard colors $> 0.1$ in Figure~\ref{fig:cc} reflect
either spectra with photon indices $\Gamma < 1$ (e.g. 
Figure~\ref{fig:spec}{\it~c}), or in rarer instances sources with strong line 
emission (Figure~\ref{fig:spec}{\it~d}). 

\section{Discussion\label{sec:dis}}

This 590~ks exposure of the 17\arcmin\ by 17\arcmin\ region around \sgrastar\ 
contains the largest number of sources ever detected in a single
field by an X-ray instrument. This sample is an order of magnitude larger
than previous X-ray surveys of the Galactic plane with \einstein\
\citep{hg84}, \rosat\ \citep{mot97, mor01}, 
and \asca\ \citep{sug01}. 
Previous observations of the Galactic center with other X-ray telescopes
only revealed 17 sources
in this field \citep{wat81,pav94,pt94,sid99,sbm01,sak02}, whereas we have 
resolved over 2000 sources.

\subsection{Numbers of Sources\label{sec:dis:num}}

The surface density of sources observed in this field is extremely 
high. Above our completeness limit of $3\times10^{-15}$ \ergcms, we 
estimate that there are approximately 15,000 sources deg$^{-2}$ in the
2.0--8.0~keV energy band in the inner 9\arcmin\ toward the Galactic
center (Equations~\ref{eq:rad} and \ref{eq:mod}, and 
Figure~\ref{fig:norm_dnds}). 
This surface density of sources is much higher than that of 
extra-galactic sources observed in fields far from the Galactic plane 
\citep{bra01,ros02}. Using the $\log(N) - \log(S)$ distribution of
\citet{bra01} and accounting conservatively for the 30\% reduction in the flux 
from background sources due to the $>5\times10^{22}$~cm$^{-2}$ of absorption 
toward the Galactic center, we would expect 630 extra-galactic sources 
deg$^{-2}$ above our completeness limit. This implies that only 4\% of the
absorbed sources brighter than $3\times10^{-15}$ \ergcms\ should be AGN. 
Applying the same computation at our detection limit of 
$4\times10^{-16}$ \ergcms, we find that only 1\% of the sources should be AGN. 
Thus, we estimate that 
between 20--100 of the 2076 sources detected only above 2~keV are background 
AGN. The density of hard 
sources toward the Galactic center is also about 50 times higher than that 
inferred by \citet{ebi02} in a field with similarly high absorption
($6\times 10^{22}$ cm$^{-2}$ of H) at
$(l, b) \approx (+28^\circ.45, 0^\circ.2)$, down to 
the same flux limit. 
 
We find that the spatial density of Galactic center X-ray sources increases 
as $R^{-2.0\pm0.1}$ approaching \sgrastar\ (assuming spherical symmetry; 
see Equation~\ref{eq:rad} and Figure~\ref{fig:radialprofile}). 
The density of stellar sources observed in the infrared also 
increases dramatically
in the inner 300 pc of our Galaxy, which is referred to as the nuclear bulge
\citep{mdz96}. Within the inner 30 pc of the nuclear bulge, the infrared 
population increases in space density approximately as $R^{-2.0\pm0.3}$, 
where $R$ is the distance from the Galactic center \citep{sm96}. 
Thus, the spatial distributions of X-ray and infrared sources are quite 
similar. This 
implies that these X-ray sources lie primarily in the nuclear bulge, and 
that their spatial distribution traces that of infrared stars.


Foreground sources (detected below 1.5~keV) have a mean surface 
density of 1800 sources deg$^{-2}$ above $3\times10^{-16}$ \ergcms\ 
(0.5--2~keV; Figure~\ref{fig:norm_dnds}). 
These soft sources are distributed uniformly over the field
(Figure~\ref{fig:spatial}), as would be expected if they lie in the 
Galactic disk. However, the density of soft sources in this field is lower 
than that of \citet{ebi02}, who detected 183 soft sources down to a limiting
flux of $7\times10^{-16}$ \ergcms\ in a 250 arcmin$^{2}$ field at 
$l \approx +28^\circ.45$. This density, 2600 sources 
deg$^{-2}$, is over 3 times higher than the density we would predict 
by extrapolating our number--flux counts to $7\times10^{-16}$ \ergcms,
770 sources deg$^{-2}$ (Equation~\ref{eq:modsoft}). The lower number of 
sources observed in the field toward the Galactic center is probably due 
to the higher absorption, which prevents us from seeing to as large a 
distance in the disk. On the other hand, the slope of our $\log(N) - \log(S)$
distribution for the foreground sources ($\alpha = 1.00 \pm 0.09$) is 
similar to those derived from 
shallower ROSAT surveys of the Galactic plane at 30--300 times higher fluxes 
($\alpha = 1.05 \pm 0.13$ in Motch \etal\ 1997; $\alpha = 1.5^{+0.7}_{-0.4}$
in Morley \etal\ 2001). 

\subsection{Point Source Contribution to the Diffuse Emission\label{sec:dis:dif}}

The point sources observed in the field contribute a significant fraction
to the X-ray emission that has previously been ascribed to diffuse
emission at the Galactic center \citep{koy96, sm99}. The flux produced by
these point sources can be estimated from the $\log(N) - \log(S)$ distribution
(Equation~\ref{eq:mod}) according to
\begin{equation}
F(> S_{\rm min}) = \int_{S_{\rm min}}^{S_{\rm max}} {{dN} \over {dS}} S dS.
\label{eq:fcont}
\end{equation}
For a power-law number count distribution, this integrates to 
\begin{equation} 
F(> S_{\rm min}) = N_o S_o {{\alpha} \over {\alpha-1}} \left[
\left( {{S_{\rm min}} \over {S_o}} \right)^{-(\alpha - 1)} - 
\left( {{S_{\rm max}} \over {S_o}} \right)^{-(\alpha - 1)} \right].
\end{equation}
We converted the photon fluxes used in Equation~\ref{eq:mod} into energy 
fluxes by assuming 
1~\phcms~$= 8\times10^{-9}$~\ergcms\ (2.0--8~keV), and integrated
over the fluxes for which our survey is complete (Table~\ref{tab:surveys}).
We find that point sources 
with fluxes greater than $3 \times 10^{-15}$ \ergcms\ contribute a mean 
surface brightness of  
$4\times10^{-14}$ \ergcms\ arcmin$^{-2}$ over the inner 9\arcmin\ around
\sgrastar. This is about 10\% of the diffuse flux from the 
inner regions of the Galaxy derived by Koyama \etal\ (1996; $10^{-9}$ \ergcms\
in a 1\degree\ field, or $3\times10^{-13}$ \ergcms\ arcmin$^{-2}$) and 
by Sidoli \& Mereghetti (2001; $1 \times 10^{-10}$ \ergcms\ in a 
190 arcmin$^{2}$ field, or $5\times10^{-13}$ \ergcms\ arcmin$^2$). 
A similar result was obtained by \citet[][]{ebi01} in a region at
$l = 28^\circ$ and $b = 0.2^{\circ}$.

However, the steep slope of the flux distribution ($\alpha = 1.7$), 
implies that the integrated flux from point sources in the field will
diverge if the distribution extends to arbitrarily low fluxes 
(see Figure~\ref{fig:norm_dnds}).
Point sources would account for all of the diffuse emission reported
by \citet{koy96} and \citet{sm99} if the luminosity distribution in 
Equation~\ref{eq:mod}
extends a factor of 40--100 lower in flux. However, from the image in 
Figure~\ref{fig:image}, it is clear that filamentary features contribute
a significant fraction of the diffuse emission, which implies that the 
flux distribution in the 2--8~keV band (where most of the diffuse emission
is observed) must turn over between fluxes of $3 \times 10^{-17}$ and 
$3 \times 10^{-15}$ \ergcms, or luminosities of $2\times 10^{29}$ to
$2\times 10^{31}$ \ergsec\ at the Galactic center. 

\subsection{The Nature of the Galactic Center Point Sources\label{sec:dis:nat}}

Sources are detected at the Galactic center 
with fluxes between $3 \times 10^{-16}$ and $2 \times 10^{-13}$ \ergcms\
(2.0--8.0~keV; compare Figure~\ref{fig:lumdist}). This translates to 
luminosities of $3\times 10^{30}$ to $2 \times 10^{33}$ \ergsec\ at 8 kpc, 
if an average decrease in flux of 30\% due to absorption is accounted
for. Comparing these values with Table~\ref{tab:ps}, only main sequence
stars later than type O have X-ray luminosities below the lower
limit of our sample. There are likely to be many examples of the other 
classes of sources in the field. Among ordinary stellar systems, 
about 10\% of the X-ray stars in the Orion Nebula Cluster \citep{fei02b}
and 70\% of RS CVn systems \citep{dem93} lie within the luminosity 
range of the detected sources. 
However, above our completeness limit of $3\times10^{31}$~\ergsec,
we should find only 1\% of YSOs and 10\% of RS CVns, and the large absorbing
column toward the Galactic center would attenuate the soft spectra of these
sources by factors of 2--6. 
These should contribute significantly only to the numbers of relatively faint 
sources with hard colors less than 0.1 (see Figures~\ref{fig:hr} 
and \ref{fig:cc}). 
Several isolated Wolf-Rayet and O stars
also have been detected at the Galactic center \citep{cot99}. Such systems 
could account for the brighter X-ray sources with soft spectra,
particularly if they are binaries.
 
With significantly higher luminosities and relatively harder spectra, 
stellar remnants probably comprise the bulk of the Galactic 
center sources. Among the sources in Table~\ref{tab:ps}, low-mass X-ray 
binaries (LMXBs) and young, isolated pulsars are certainly luminous enough 
to be seen at the Galactic center, but they are expected to be quite rare,
with only on order $10^3$ in the entire Galaxy \citep{itf97,rom98,pos02}. 
Few examples of these should be present in our sample. On the other hand, 
Howell, Nelson, \& Rappaport (2001)\nocite{hnr01} suggest that 
there could be $10^6$ cataclysmic variables (CVs) in our Galaxy. We
would expect to detect 60\% of CVs containing 
weakly-magnetized white dwarfs above
our detection limit, and 7\% above our completeness limit \citep{ver97}.
However, most of these sources CVs (namely, the un-magnetized ones) 
also have relatively soft, thermal spectra (Table~\ref{tab:ps}).

The best candidates for the very hard X-ray sources in Figure~\ref{fig:cc} 
are CVs
containing magnetized white dwarfs (polars and intermediate polars) and
neutron stars in HMXBs. 
Magnetized CVs comprise on order 10\% of accreting white dwarfs 
\citep[e.g.][]{war95}, and often exhibit flat spectra that are 
typically interpreted as hot ($kT > 10$~keV) thermal spectra with multiple 
absorbing components that partially cover the X-ray emitting region 
\citep[e.g.,][]{ei99, sug00}.
Wind-accreting X-ray pulsars in HMXBs also frequently 
exhibit non-thermal spectra that can be described by $\Gamma \sim 0$ 
power laws below 10~keV \citep[e.g.]{cam01}. 
Although HMXBs are usually observed with luminosities 
above $10^{33}$ \ergsec\ \citep{cam02}, several faint, pulsing sources 
recently have been discovered that are thought to be accreting neutron 
stars \citep{kin98, tor99, oos99, sak00}. 
Pfahl, Rappaport, \& Podsiadlowski (2002)\nocite{pfa02} predict that there 
could be a few times $10^4$ wind-accreting neutron stars (HXMBs) in the 
Milky Way. 
Hard sources similar to those found in our survey are almost certainly
fairly common in the Galactic disk, as an ASCA survey of the central region 
of the Galactic plane ($|l| < 45^\circ, |b| < 0^\circ.5$) turned up about a
dozen sources with $\Gamma < 1$ power laws out of 163 sources  with fluxes 
as low as $10^{-12.5}$ \ergcms.

Whatever the nature of these hard sources, they represent an 
unprecedented sample of X-ray emitting objects. 
More than half of the sources brighter than $4\times10^{-15}$~\ergcms\ 
(2.0--8.0~keV) for which we can derive medium and hard colors have spectra 
consistent with $\Gamma < 1$ power laws (Figure~\ref{fig:cc}), which implies 
that there could be over 1000 of these hard sources
in the inner 20~pc around \sgrastar. In contrast, only on 
order 100 magnetic CVs
\citep{war95, ei99} and wind-accreting pulsars (Liu, van Paradijs, 
 \& van den Heuvel 2000)\nocite{liu00} are known in our Galaxy.

In fact, it may prove difficult to understand why so many magnetized CVs
or neutron star HMXBs would be present at the Galactic center.  
Assuming that all of the point sources are members of the nuclear bulge
(which is about 300 pc across in the radial direction; Mezger \etal\ 1996), 
our image surveys a volume of approximately 
$40 \cdot 40 \cdot 300 = 5\times10^5$~pc$^3$,
and the average stellar number density of the field is on order 1000 times
that in the local neighborhood \citep[compare][Launhardt, Zylka, \&
 Mezger 2002]{bm98}\nocite{lzm02}. If we take 
the local the local number density of polars and intermediate polars to 
be $3\times10^{-7}$ pc$^{-3}$ \citep{war95}, and assume that their 
density can be scaled to the Galactic center according to the stellar density, 
we would expect to observe only $\sim 150$ in our current image. 
Unfortunately, this estimate is only reliable to within an order of
magnitude, because various selection effects in surveys of CVs introduce 
large uncertainties into ({\it i}) estimates of the local space density of 
CVs \citep[compare, e.g.,][]{war95,sch02} and ({\it ii}) the fraction of 
magnetic CVs thought to exhibit flat spectra between 2--8~keV 
\citep[compare][Haberl, Motch, \& Zickgraf 2002]{ver97,ei99}\nocite{hmz02}.

Since the nuclear bulge contains on order 1\% of the mass of the 
Galactic disk \citep{mdz96}, Pfahl \etal\ (2002) have predicted that 
it could contain several hundred of the young, wind-accreting neutron 
stars that may exist in the Galaxy. The early-type 
mass donors in these systems only have a typical lifetime of 
$5\times10^{7}$ years \citep{pfa02}, so the main theoretical uncertainty 
in determining the number of HXMBs expected in this field is the 
rate of recent star formation in the nuclear bulge \citep{mdz96,sm96}.
However, the large numbers of faint neutron star HMXBs that
\citet{pfa02} predict have not yet
been identified in the Galaxy --- the $\sim 100$ known wind-accreting
neutron stars are mostly transient systems with observed luminosities 
that range from 
$10^{35} - 10^{38}$ \ergsec. The one system that has been 
detected with $L_{\rm X} < 10^{33}$~\ergsec\ had a soft spectrum 
that was consistent with a $\Gamma \sim 2.5$ power law 
\citep[V~0332$+$53; see][]{cam02}, and would not be among the 
hard sample from our image if it were 
placed at the Galactic Center (compare Figure~\ref{fig:cc}).

\section{Conclusions\label{sec:conc}}
We have presented a sample of 2357 X-ray sources detected during 590~ks 
of \chandra\ observations of the 17\arcmin\ by 17\arcmin\ field around 
\sgrastar\ (Figures~\ref{fig:rawimage} and \ref{fig:image}). 
The completeness limit of our survey at the Galactic center is about 
$3\times10^{-15}$ \ergcms (2--8~keV), 
while sources are detected with fluxes nearly an order of magnitude lower
(Figure~\ref{fig:lumdist}).
Only 20--100 of these sources are expected to be background AGN.
The large number of sources in this field probably results from the high 
stellar density at the Galactic center. Indeed, we have demonstrated that
the surface density of Galactic center X-ray sources decreases as $1/\theta$
away from \sgrastar\ (Figure~\ref{fig:radialprofile}), just as the surface 
density of infrared stars does
\citep{sm96}. We have also shown that the $\log(N)-\log(S)$ distribution
of the Galactic center sources is very steep, rising as $S^{-1.7}$ near our
completeness limit (Figure~\ref{fig:norm_dnds}). 
This indicates that point
sources can contribute significantly to the diffuse component of the 
Galactic X-ray emission.

More than half of the sources for which we have spectral information are
very hard, with spectra that are consistent with $\Gamma < 1$ power laws
(Figure~\ref{fig:cc}).
Such hard spectra have only been observed previously from magnetically 
accreting white dwarfs and wind-accreting neutron stars. It these X-ray sources
are magnetic CVs, they would be the first low-mass stars identified in the 
nuclear bulge. If they are wind-accreting neutron stars, these systems 
would provide an important constraint on the amount of star formation that
has taken place near the Galactic center in the last $10^7-10^8$ years. 
This highlights the importance of identifying the nature of the Galactic
center sources with more certainty. The X-ray spectral and timing properties 
of these sources will be reported in detail in the near future, and 
we are in the process of identifying these sources at radio and infrared
wavelengths. 

\acknowledgements{
We gratefully thank D. Schwartz, P. Slane, and the \chandra\ Mission Planning
group for their efforts in scheduling the observations in late May so 
that they would have nearly identical roll angles and aim points.
We also thank K. Getman and F. Bauer for developing software that aided us
in collating and visualizing these results, V. te Velde for helping to 
construct a map of the diffuse background, P. Schechter for valuable 
advice on how to treat the $\log(N)-\log(S)$ distribution, 
and E. Pfahl and J. Sokoloski for helpful discussions about the possible 
natures of these sources. We also thank the referee, P. Edmonds, for his
careful review of the manuscript. This work has been supported by NASA grants
NAS 8-39073 and NAS 8-00128. W.N.B. also acknowledges the LTSA grant 
NAG 5-8107 and the Alfred P. Sloan foundation.}

\clearpage

\begin{deluxetable}{lccc}
\tablecolumns{4}
\tablewidth{0pc}
\tablecaption{Galactic X-ray Point Sources\label{tab:ps}} 
\tablehead{
\colhead{Object} & \colhead{$\log(L_{\rm X})$\tablenotemark{a}} 
& \colhead{Spectrum\tablenotemark{b}} & 
\colhead{References\tablenotemark{c}} \\
\colhead{} & \colhead{$\log(\rm{erg~s}^{-1})$} & \colhead{} & \colhead{}
} \startdata
MS Stars\tablenotemark{d} & $25-30.3$ & $kT<1$~keV Plasma & [1,2] \\
YSOs & $29-31.1$ & $kT = 1-10$~keV Plasma & [3,4,5] \\
RS CVn/Algol & $29-31.7$ & $kT = 0.1-2$~keV Plasma  & [6,7] \\
WR/O Stars & $31-35$ & $kT = 0.1-6$~keV Plasma & [8,9,10] \\
CVs & $29.5-32.6$ & $kT = 1-25$~keV Plasma & [11,12,13,14] \\
Pulsars & $29.3-39$ & $\Gamma = 1-2.5$ PL ; $kT = 0.3$~keV BB & [15,16] \\
NS LMXBs & $31.6-38$ & $kT\sim0.3$~keV BB ; $\Gamma=1-2$ PL\tablenotemark{e} & [17,18,19,20] \\
BH LMXBs & $30-39$ & $\Gamma=1-2$ PL\tablenotemark{e} & [17,21] \\
HMXBs & $32.7-38$ & $\Gamma = 0.5-2.5$ PL & [22,23] \\
\enddata
\tablenotetext{a}{Luminosities represent ranges reported in the literature. 
Below $L_{\rm X} \sim 10^{29}$ \ergsec, sources are difficult to detect, 
and lower bounds at this level generally represent the sensitivity limits of 
the respective observations.}
\tablenotetext{b}{Spectra of point sources are typically
described by thermal plasma \citep{rs77,mew86}, power 
laws (denoted by PL), or blackbodies (denoted by BB).}
\tablenotetext{c}{The references 
are not a complete compilation, but represent a sampling of surveys
and recent results that are amenable to comparisons with observations
in the \chandra\ bandpass (0.5--10~keV).}
\tablenotetext{d}{Later than type O.}
\tablenotetext{e}{For the LMXBs,
we include only spectral properties in quiescence ($L_{\rm X} < 10^{34}$
\ergsec).}
\tablerefs{[1] \citet{kri01}; [2] \citet{hem95}; [3] \citet{gar00}; 
[4] \citet{pz02}; [5] Kohno, Koyama, \& Hamaguchi (2002)\nocite{koh02}; 
[6] Singh, Drake, \& White (1996)\nocite{sdw96}; [7] \citet{dem93};
[8] \citet{yz02}; [9] Portegies-Zwart, Pooley, \& Lewin (2002)\nocite{poz02}; 
[10] \citet{pol87}; [11] \citet{ver97};
[12] \citet{muk00}; [13] \citet{mm02}; [14] \citet{szk02}; 
[15] \citet{ba02}; [16] \citet{pos02}; [17] \citet{asa98}; 
[18] \citet{rut01}; [19] \citet{wij02}; [20] \citet{cam02}; 
[21] \citet{kon02}; [22] \citet{cam01}; [23] \citet{cam02b}}
\end{deluxetable}

\begin{deluxetable}{lccccc}
\tablecolumns{6}
\tablewidth{0pc}
\tablecaption{Observations of the Inner 20 pc of the Galaxy\label{tab:obs}}
\tablehead{
\colhead{} & \colhead{} & \colhead{} & 
\multicolumn{2}{c}{Aim Point} & \colhead{} \\
\colhead{Start Time} & \colhead{Sequence} & \colhead{Exposure} & 
\colhead{RA} & \colhead{DEC} & \colhead{Roll} \\
\colhead{(UT)} & \colhead{} & \colhead{(s)} 
& \multicolumn{2}{c}{(degrees J2000)} & \colhead{(degrees)}
} \startdata
1999 Sep 21 02:43:00 & 0242  & 40,872 & 266.41382 & -29.0130 & 268 \\
2000 Oct 26 18:15:11 & 1561 & 35,705 & 266.41344 & -29.0128 & 265 \\
2001 Jul 14 01:51:10 & 1561 & 13,504 & 266.41344 & -29.0128 & 265 \\
2002 Feb 19 14:27:32 & 2951  & 12,370 & 266.41867 & -29.0033 & 91 \\
2002 Mar 23 12:25:04 & 2952  & 11,859 & 266.41897 & -29.0034 & 88 \\
2002 Apr 19 10:39:01 & 2953  & 11,632 & 266.41923 & -29.0034 & 85 \\
2002 May 07 09:25:07 & 2954  & 12,455 & 266.41938 & -29.0037 & 82 \\
2002 May 22 22:59:15 & 2943  & 34,651 & 266.41991 & -29.0041 & 76 \\
2002 May 24 11:50:13 & 3663  & 37,959 & 266.41993 & -29.0041 & 76 \\
2002 May 25 15:16:03 & 3392  & 166,690 & 266.41992 & -29.0041 & 76 \\
2002 May 28 05:34:44 & 3393  & 158,026 & 266.41992 & -29.0041 & 76 \\
2002 Jun 03 01:24:37 & 3665  & 89,928 & 266.41992 & -29.0041 & 76 \\
\enddata
\end{deluxetable}

\begin{deluxetable}{lcccccccccccccc}
\rotate
\tabletypesize{\tiny}
\tablecolumns{15}
\tablewidth{0pc}
\tablecaption{Brightest Point Sources in the 17\arcmin\ by 17\arcmin\ Field 
toward the Galactic Center\label{tab:cat}}
\tablehead{
\colhead{Source Name} & \colhead{RA} & \colhead{DEC} & 
\colhead{Offset} & \colhead{$T_{\rm live}$} & 
\colhead{$f_{\rm PSF}$} & \colhead{$E_{\rm PSF}$} &
\colhead{Net Cts} & \colhead{Soft} & \colhead{Med} &\colhead{Hard} & 
\multicolumn{4}{c}{Fluxes ($10^{-7}$~\phcms)} \\
\colhead{} & \multicolumn{2}{c}{(J2000)} & 
\colhead{(\arcmin)} & \colhead{(ks)} & \colhead{} & \colhead{(keV)} &
\colhead{$F$} & 
\colhead{Color} & \colhead{Color} & \colhead{Color} & 
\colhead{$S$} & \colhead{$M1$}
& \colhead{$M2$} & \colhead{$H$} \\
\colhead{(1)} & \colhead{(2)} & \colhead{(3)} &
\colhead{(4)} & \colhead{(5)} & \colhead{(6)} & \colhead{(7)} &
\colhead{(24--26)} & 
\colhead{(27--29)} & \colhead{(30--32)} & \colhead{(33--35)} & 
\colhead{(36--38)} & \colhead{(39--41)} 
& \colhead{(42--44)} & \colhead{(45--47)} 
 } 
\startdata
174530.0$-$290704 & 266.37505 & $-$29.11780 &  7.0 & 625.6 & 0.90 & 1.5 & $ 5585(77) $ & $-0.975_{- 0.028}^{+ 0.027}$ & $-0.743_{- 0.407}^{+ 0.302}$ & $-1.000^{+ 1.187}$& $ 255(4) $ & $ 4.5(8) $ & $ 0.6(5) $ & $  < 1.73$\\
174541.0$-$290014 & 266.42083 & $-$29.00399 &  0.3 & 625.6 & 0.87 & 4.5 & $ 2916(57) $ & $ 0.769_{- 0.030}^{+ 0.024}$ & $ 0.767_{- 0.008}^{+ 0.008}$ & $ 0.275_{- 0.015}^{+ 0.014}$& $ 0.7(3) $ & $ 7.4(9) $ & $ 47(2) $ & $ 162(4) $\\
174536.1$-$285638 & 266.40059 & $-$28.94407 &  3.9 & 625.6 & 0.90 & 1.5 & $ 2818(54) $ & $ 0.797_{- 0.007}^{+ 0.006}$ & $ 0.065_{- 0.022}^{+ 0.021}$ & $-0.186_{- 0.030}^{+ 0.029}$& $ 4.6(5) $ & $ 56(2) $ & $ 53(2) $ & $ 75(3) $\\
174607.5$-$285951 & 266.53132 & $-$28.99757 &  6.0 & 625.6 & 0.90 & 1.5 & $ 2670(54) $ & $-0.355_{- 0.033}^{+ 0.031}$ & $-0.215_{- 0.042}^{+ 0.040}$ & $-0.250_{- 0.055}^{+ 0.052}$& $ 62(2) $ & $ 40(2) $ & $ 22(1) $ & $ 27(2) $\\
174543.9$-$290456 & 266.43305 & $-$29.08238 &  4.6 & 625.6 & 0.90 & 1.5 & $ 1936(45) $ & $-0.957_{- 0.047}^{+ 0.045}$ & $-0.341_{- 0.266}^{+ 0.207}$ & $-1.000^{+ 0.762}$& $ 83(2) $ & $ 2.5(5) $ & $ 1.0(3) $ & $ < 1.28$\\ [5pt]
174541.5$-$285814 & 266.42296 & $-$28.97080 &  2.2 & 625.6 & 0.90 & 1.5 & $ 1839(44) $ & $ 0.579_{- 0.020}^{+ 0.019}$ & $ 0.201_{- 0.026}^{+ 0.025}$ & $ 0.006_{- 0.030}^{+ 0.028}$& $ 4.7(5) $ & $ 24(1) $ & $ 31(1) $ & $ 63(3) $\\
174552.2$-$290744 & 266.46754 & $-$29.12908 &  7.7 & 625.6 & 0.74 & 4.5 & $ 1466(40) $ & $-0.833_{- 0.053}^{+ 0.050}$ & $-0.494_{- 0.165}^{+ 0.142}$ & $-0.480_{- 0.386}^{+ 0.281}$& $ 72(2) $ & $ 9(1) $ & $ 2.5(6) $ & $ 2(1) $\\
174532.7$-$290552 & 266.38663 & $-$29.09785 &  5.6 & 625.6 & 0.90 & 1.5 & $ 1379(39) $ & $-0.262_{- 0.045}^{+ 0.043}$ & $-0.139_{- 0.052}^{+ 0.049}$ & $-0.263_{- 0.072}^{+ 0.067}$& $ 27(1) $ & $ 22(1) $ & $ 13.6(9) $ & $ 17(2) $\\
174558.9$-$290724 & 266.49557 & $-$29.12340 &  8.1 & 625.6 & 0.90 & 4.5 & $ 1346(41) $ & $ 1.000_{- 0.192}$ & $ 0.624_{- 0.020}^{+ 0.018}$ & $ 0.301_{- 0.022}^{+ 0.021}$& $ <0.50$ & $ 6.4(9) $ & $ 23(1) $ & $ 94(4) $\\
174539.7$-$290029 & 266.41567 & $-$29.00827 &  0.1 & 625.6 & 0.90 & 4.5 & $ 1229(61) $ & $ 0.909_{- 0.009}^{+ 0.008}$ & $ 0.256_{- 0.042}^{+ 0.039}$ & $-0.410_{- 0.088}^{+ 0.081}$& $ 0.7(4) $ & $ 20(2) $ & $ 28(2) $ & $ 23(3) $\\ [5pt]
174538.0$-$290022 & 266.40861 & $-$29.00619 &  0.4 & 625.6 & 0.87 & 4.5 & $ 1131(36) $ & $ 0.913_{- 0.007}^{+ 0.006}$ & $ 0.349_{- 0.026}^{+ 0.025}$ & $-0.254_{- 0.048}^{+ 0.046}$& $ 0.5(2) $ & $ 15(1) $ & $ 26(1) $ & $ 30(2) $\\
174548.9$-$285751 & 266.45400 & $-$28.96439 &  3.3 & 625.6 & 0.90 & 4.5 & $ 1127(35) $ & $ 1.000_{- 0.163}$ & $ 0.566_{- 0.022}^{+ 0.020}$ & $ 0.213_{- 0.027}^{+ 0.026}$& $  <0.40$ & $ 6.2(7) $ & $ 19(1) $ & $ 58(2) $\\
174550.5$-$285239 & 266.46067 & $-$28.87773 &  8.1 & 584.8 & 0.90 & 1.5 & $ 1088(38) $ & $-0.921_{- 0.067}^{+ 0.063}$ & $-0.315_{- 0.421}^{+ 0.289}$ & $-1.000^{+ 0.831}$& $ 60(2) $ & $ 3(1) $ & $ 1.5(7) $ & $< 2.25$\\
174545.2$-$285828 & 266.43870 & $-$28.97466 &  2.3 & 625.6 & 0.81 & 1.5 & $ 1027(33) $ & $-0.865_{- 0.064}^{+ 0.061}$ & $-0.381_{- 0.190}^{+ 0.158}$ & $-0.150_{- 0.248}^{+ 0.190}$& $ 71(2) $ & $ 7(1) $ & $ 2.7(7) $ & $ 4(1) $\\
174535.5$-$290124 & 266.39823 & $-$29.02336 &  1.4 & 625.6 & 0.90 & 4.5 & $ 1011(33) $ & $ 1.000_{- 0.132}$ & $ 0.694_{- 0.017}^{+ 0.015}$ & $ 0.199_{- 0.029}^{+ 0.027}$& $ <0.20$ & $ 3.8(6) $ & $ 18(1) $ & $ 52(2) $\\ [5pt]
174520.6$-$290152 & 266.33585 & $-$29.03113 &  4.5 & 625.6 & 0.88 & 1.5 & $ 1008(33) $ & $-0.660_{- 0.059}^{+ 0.056}$ & $-0.459_{- 0.120}^{+ 0.107}$ & $-0.351_{- 0.202}^{+ 0.166}$& $ 40(1) $ & $ 12(1) $ & $ 3.4(5) $ & $ 3.4(9) $\\
174540.1$-$290055 & 266.41733 & $-$29.01546 &  0.5 & 625.6 & 0.90 & 4.5 & $ 954(36) $ & $ 0.862_{- 0.021}^{+ 0.016}$ & $ 0.564_{- 0.025}^{+ 0.023}$ & $ 0.037_{- 0.040}^{+ 0.038}$& $ 0.4(2) $ & $ 6.5(9) $ & $ 19(1) $ & $ 41(2) $\\
174534.5$-$290201 & 266.39408 & $-$29.03363 &  2.0 & 625.6 & 0.90 & 4.5 & $ 895(31) $ & $ 0.734_{- 0.081}^{+ 0.055}$ & $ 0.769_{- 0.017}^{+ 0.015}$ & $ 0.474_{- 0.020}^{+ 0.019}$& $ 0.2(2) $ & $ 1.6(4) $ & $ 10.4(8) $ & $ 58(2) $\\
174545.5$-$285829 & 266.43972 & $-$28.97475 &  2.3 & 625.6 & 0.80 & 4.5 & $ 876(30) $ & $ 0.847_{- 0.025}^{+ 0.020}$ & $ 0.661_{- 0.020}^{+ 0.018}$ & $ 0.199_{- 0.031}^{+ 0.029}$& $ 0.6(4) $ & $ 9(1) $ & $ 39(2) $ & $ 118(6) $\\
174544.9$-$290027 & 266.43745 & $-$29.00759 &  1.1 & 625.6 & 0.72 & 4.5 & $ 839(31) $ & $ 0.745_{- 0.028}^{+ 0.024}$ & $ 0.396_{- 0.036}^{+ 0.033}$ & $ 0.223_{- 0.033}^{+ 0.031}$& $ 0.9(3) $ & $ 8.5(9) $ & $ 17(1) $ & $ 53(3) $\\ [5pt]
174552.0$-$285312 & 266.46681 & $-$28.88676 &  7.7 & 625.6 & 0.90 & 4.5 & $ 817(33) $ & $ 0.677_{- 0.080}^{+ 0.058}$ & $ 0.619_{- 0.030}^{+ 0.027}$ & $ 0.351_{- 0.028}^{+ 0.026}$& $ 1.0(7) $ & $ 7(1) $ & $ 25(2) $ & $ 111(6) $\\
174513.1$-$285624 & 266.30498 & $-$28.94007 &  7.1 & 625.6 & 0.90 & 4.5 & $ 788(32) $ & $ 0.527_{- 0.112}^{+ 0.083}$ & $ 0.671_{- 0.024}^{+ 0.021}$ & $ 0.218_{- 0.034}^{+ 0.032}$& $ 0.8(4) $ & $ 3.4(8) $ & $ 15(1) $ & $ 49(3) $\\
174549.3$-$285557 & 266.45546 & $-$28.93276 &  4.9 & 625.6 & 0.90 & 4.5 & $ 782(30) $ & $ 0.775_{- 0.038}^{+ 0.030}$ & $ 0.510_{- 0.034}^{+ 0.031}$ & $ 0.315_{- 0.029}^{+ 0.028}$& $ 0.5(3) $ & $ 5.5(8) $ & $ 14(1) $ & $ 56(3) $\\
174547.0$-$285333 & 266.44610 & $-$28.89252 &  7.1 & 625.6 & 0.90 & 4.5 & $ 778(33) $ & $ 1.000_{- 0.474}$ & $ 0.729_{- 0.021}^{+ 0.019}$ & $ 0.272_{- 0.032}^{+ 0.030}$& $  <0.59$ & $ 2.6(7) $ & $ 14(1) $ & $ 52(3) $\\
174527.6$-$285258 & 266.36518 & $-$28.88304 &  8.0 & 625.6 & 0.90 & 4.5 & $ 758(32) $ & $ 1.000_{- 0.853}$ & $ 0.821_{- 0.017}^{+ 0.015}$ & $ 0.437_{- 0.025}^{+ 0.023}$& $  <0.71$ & $ 1.3(6) $ & $ 11(1) $ & $ 62(3) $\\
\enddata
\tablecomments{This is a portion of the full table, which is available via
the electron version of this paper. The columns are as follows: 
(1) Source name, which should be appended to the IAU designation CXOGC~J. 
(2-3) The ra and dec in decimal degrees, J2000.
(4) Offset of source from nominal aim point, \sgrastar (see text). 
(5) The sum of the live times for all of the observations in which 
a source was detected, used to compute photon fluxes. Note that the 
effective exposure factoring in proximity to chip gaps and bad columns is 
encoded in the mean value of the ARF. 
(6) Fraction of the PSF energy enclosed by source extraction region. Sources
with fractions near 70\% are likely to be confused.
(7) Energy at which PSF was estimated for $f_{\rm PSF}$, 
where $E_{\rm PSF} = 1.5$ keV for foreground sources and 
$E_{\rm PSF} = 4.5$ keV for sources at or beyond the galactic center. 
(8-23) In the machine-readable version of this table we list for each of the 
four energy bands: the total counts 
extracted from the source region, the estimated background in the 
source region, the ratio of the areas of the background and source regions 
weighted by the number of background counts, and the mean value of the 
ARF. (24-26) Net counts in the full band $F = $ 0.5--8.0~keV.
(27-35) Colors are defined 
according to $(h-s)/(h+s)$, where $h$ and $s$ are the net
counts in high and low energy bands, respectively. For the soft color,
$h$ is in the 2.0--3.3 keV band, and $s$ is in the the 0.5--2.0 keV band.  
For the medium color,
$h$ is in the 3.3--4.7 keV band, and $s$ is in the the 2.0--3.3 keV band.  
For the
hard color, $h$ is 4.7--8.0 keV and $s$ is 3.3--4.7 keV. (35-47) Photon
fluxes with 1-$\sigma$ uncertainties on the last significant figure in 
parenthesis, or 90\% upper limits in each of four energy bands: 
$S = $ 0.5--2.0~keV, $M1 = $ 2.0--3.3~keV, $M2 = $ 3.3--4.7~keV, and 
$H = $ 4.7--8.0~keV.}
\end{deluxetable}

\begin{deluxetable}{cccc}
\tablecolumns{4}
\tablewidth{0pc}
\tablecaption{Limiting Fluxes for the $\log N - \log S$ Distribution
\label{tab:surveys}}
\tablehead{
\colhead{Max. Offset} & \colhead{Flux Limit} & \colhead{Solid Angle} & 
\colhead{Number} \\
\colhead{(arcmin)} & \colhead{\phcms} & \colhead{(arcmin$^2$)} &
\colhead{of Sources}
} \startdata
\cutinhead{Galactic Center Sources} 
5 & $4\times10^{-7}$ & 60.6 & 335 \\
7 & $7\times10^{-7}$ & 148.3 & 241 \\
9 & $2.8\times10^{-6}$ & 241.7 & 61 \\
\cutinhead{Foreground Sources}
7 & $1.6\times10^{-7}$ & 143.4 & 48 \\
9 & $3.2\times10^{-7}$ & 240.9 & 66 \\
\enddata
\end{deluxetable}
\end{document}